\documentclass[%
 reprint,
superscriptaddress,
footinbib,
 amsmath,amssymb,
 aps,
]{revtex4-1}

\usepackage{graphicx}
\usepackage{dcolumn}
\usepackage{bm}
\usepackage{multirow}
\usepackage{indentfirst}
\usepackage{mathtools}
\usepackage{eufrak}
\usepackage{subfigure}
\usepackage{appendix}

\begin{document}

\preprint{APS/123-QED}

\title{Helical transport in coupled resonator waveguides}

\author{JungYun Han}
\affiliation{
 Center for Theoretical Physics of Complex Systems, Institute for Basic Science (IBS), Daejeon 34126, Republic of Korea.}
 \affiliation{
 Basic Science Program, University of Science and Technology (UST), Daejeon 34113, Republic of Korea.}
 \author{Clemens Gneiting}
 \affiliation{Theoretical Quantum Physics Laboratory, RIKEN Cluster for Pioneering Research, Wako-shi, Saitama 351-0198, Japan}
\author{Daniel Leykam}
\affiliation{
 Center for Theoretical Physics of Complex Systems, Institute for Basic Science (IBS), Daejeon 34126, Republic of Korea.}
  \affiliation{
 Basic Science Program, University of Science and Technology (UST), Daejeon 34113, Republic of Korea.}
\email{dleykam@ibs.re.kr}

\date{\today}

\begin{abstract}
We show that a synthetic pseudospin-momentum coupling can be used to design quasi-one-dimensional disorder-resistant coupled resonator optical waveguides (CROW). In this structure, the propagating Bloch waves exhibit a pseudospin-momentum locking at specific momenta where backscattering is suppressed. We quantify this resistance to disorder using two methods. First, we calculate the Anderson localization length $\xi$, obtaining an order of magnitude enhancement compared to a conventional CROW for typical device parameters. Second, we study propagation in the time domain, finding that the loss of wavepacket purity in the presence of disorder rapidly saturates, indicating the preservation of phase information before the onset of Anderson localization. Our approach of directly optimizing the bulk Bloch waves is a promising alternative to disorder-robust transport based on higher dimensional topological edge states. 
\end{abstract}

\pacs{Valid PACS appear here}
\maketitle


\section{Introduction}
\begin{figure}
\centering
\includegraphics[width=\linewidth, height=3cm]{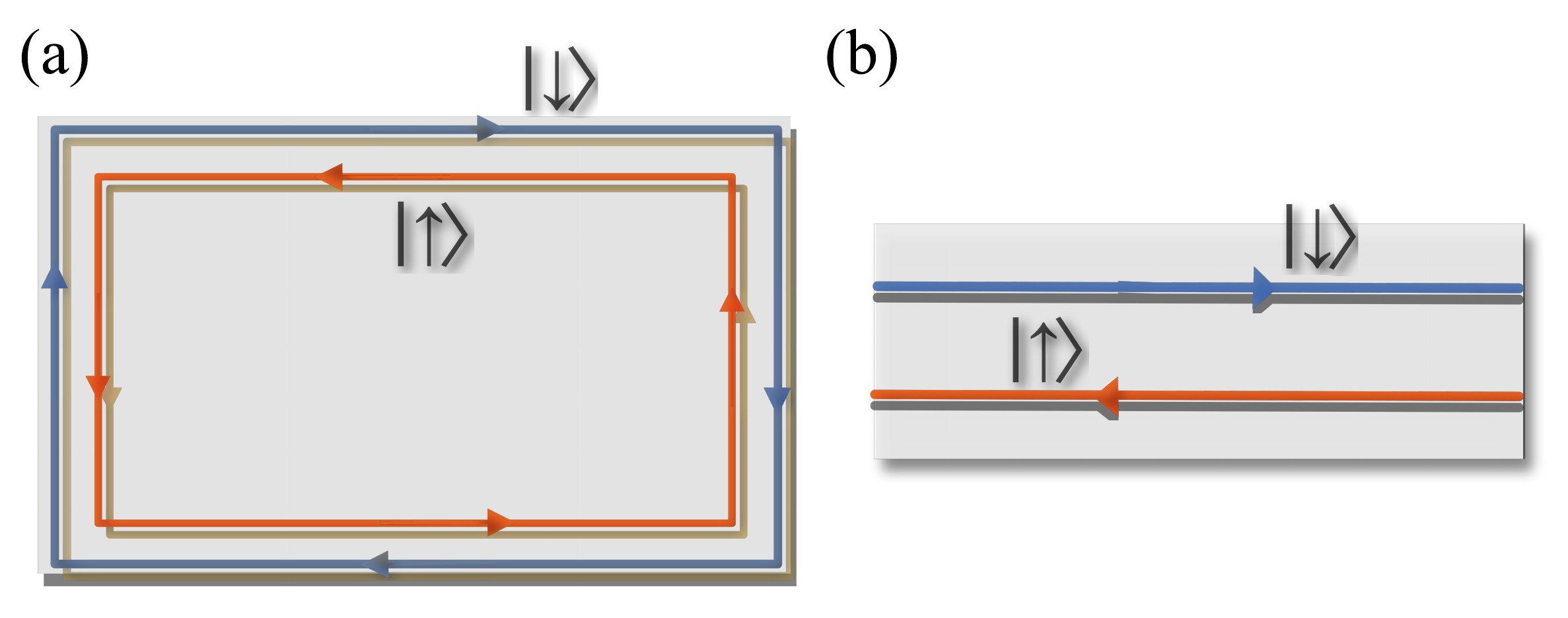}

\caption{(a) Schematic illustration of conventional helical transport along the edge of a two-dimensional topological system. (b) One-dimensional system with helical transport in its bulk dispersion relation. }
\label{helical transport}

\end{figure}

Topological phases have emerged as a powerful new paradigm for achieving disorder-robust transport in electronic condensed matter systems~\cite{Hasan10,Qi11,asboth16}. In particular, two-dimensional quantum spin Hall phases can be induced by strong spin-orbit coupling and support backscattering-immune helical edge states protected by time reversal symmetry~\cite{Kane05,Konig07}. Such helical edge states exhibit spin-momentum locking, where the propagation direction is determined by the spin, see Fig.~\ref{helical transport}(a). This spin-momentum locking can also be demonstrated for bosons if appropriate crystalline or internal symmetries replace the fermionic time reversal symmetry, for example in phononic metamaterials~\cite{Susstrunk15}, optical lattices for cold atoms \cite{Budich15, Budich17}, and photonic systems. 

Photonic topological phases were first demonstrated 10 years ago, motivated by their potential for designing disorder-robust optical waveguides. The first experiments were based on time-reversal symmetry breaking for microwaves using the magneto-optic effect~\cite{Wang08,Wang09}, and there are now many different approaches towards realizing them in time-reversal symmetric systems at optical frequencies 
\cite{Ozawa18, Khanikaev17, Lu14, Lu16b, Sun17}. Spin-momentum locking can occur where spin can be either physical spin (polarization) or some other internal degree of freedom such as sublattices or orbital angular momentum states.

One limitation of existing topologically-protected waveguide designs is that they are typically based on two or three-dimensional topological phases \cite{Hafezi11, Khanikaev12, Umucalilar11,Qian18,Leykam18}, requiring a large physical device size and increasing the cost of fabrication. To miniaturize further, it would be preferable to use one-dimensional structures. However, one-dimensional hermitian topological phases are characterized by localized end states, which by themselves do not support net transport. While approaches based on synthetic dimensions~\cite{Lustig18} or adiabatic pumping~\cite{Kraus12} are compatible with topological transport confined to a single spatial dimension, they require modulation in time, which poses an additional challenge. Approaches based on non-Hermitian delocalization are also challenging, requiring the introduction of gain and/or loss to the system~\cite{Schomerus14,Hatano96,Longhi15,Longhi15B}. 

Spin-momentum locked transport protected against certain classes of disorder is also possible in static one-dimensional systems using a combination of strong spin-orbit coupling and and an applied magnetic field. Similar to two-dimensional time-reversal symmetric topological insulators, backscattering requires a spin flip, i.e. T-breaking (magnetic) disorder, see Fig.~\ref{helical transport}(b). This helical transport with characteristic half-integer quantized conductance has been observed in one-dimensional quantum wires~\cite{Oreg10,Quay10,Kammhuber17,heedt17,Sadashige17}.  

The requisite ingredients of strong spin-orbit coupling combined with an effective magnetic field can readily be implemented in photonic systems such as coupled resonator lattices, waveguides, and microcavities~\cite{Shelykh10,Whittaker18}. For example, Ref.~\cite{Hafezi13} demonstrated an effective magnetic field for light in two-dimensional coupled resonator lattices, and spin-orbit coupling was emulated using tilted waveguide arrays arranged into a two-leg ladder~\cite{Plotnik16}. However, to the best of our knowledge, the combination of these two effects to achieve one-dimensional disorder-resistant transport has not been explored in photonics.

In this manuscript we show how to combine T-symmetry breaking with synthetic spin-orbit coupling to induce one-dimensional helical transport in coupled ring resonator optical waveguides (CROWs)~\cite{Yariv99,Cooper10,Canciamilla10,Morichetti12,Takesue13}. We show that this enables waveguiding that is less susceptible to the dominant class of disorder in this platform - disorder in the resonant frequencies of the individual resonators. We demonstrate this disorder protection both analytically and numerically using two complimentary methods. First, we calculate the scattering length under the Born approximation, obtaining an order of magnitude enhancement of the Anderson localization length around a critical energy due the spin-momentum locking. Second, we study the propagation dynamics, employing a recently-developed master equation framework ~\cite{Kropf16,Gneiting17,Gneiting18disorder} to quantify the preservation of phase information and spatial coherence of wavepackets propagating along the waveguide by calculating the purity of field, which provides the information about how much an evolving wavepacket deviates from the disorder-free state. We conclude that one-dimensional helical channels are promising way to achieve compact, disorder-resistant integrated photonic waveguides.


The rest of the paper is structured as follows: Sec. II introduces our tight binding model and discusses sources of disorder. Sec. III computes the Anderson localization length analytically and numerically, demonstrating a suppression of localization due to spin-momentum locking. Sec. IV studies the propagation of wavepackets in the time domain, showing preservation of their purity even in the presence of moderate disorder. Sec. V concludes with a summary and final remarks. Appendix A compares our results obtained under the tight binding approximation against a full scattering matrix model, demonstrating excellent agreement under typical device parameters. Appendices B and C present details of the analytical and numerical calculations of the localization length.  


\section{Model}

We will consider light propagation in an array of coupled ring resonators. Each ring hosts a set of resonant modes, with frequency spacing dictated by the free spectral range FSR = $c/ (L n_{g})$ where $c$ is the speed of light, $L$ is the length of the ring cavity, $n_{g} = n_{\text{eff}} - \lambda  \frac{dn_{\text{eff}}}{d\lambda}$ is the modal group velocity at the operating wavelength $\lambda$, and $n_{\text{eff}}$ is the effective refractive index~\cite{rabus07}. Similar to the scheme previously used in Refs.~\cite{Hafezi11,Hafezi13,Mittal16,Leykam18}, we assume clockwise and anticlockwise modes in the individual rings are decoupled, and that resonant ``site'' rings are coupled via off-resonant ``link'' rings. The former enables T-symmetry to be effectively broken via excitation of a specific mode handedness, while the latter allows tailoring of the effective spin-orbit coupling.
\begin{figure}
\centering
\includegraphics[width=\columnwidth,height=9cm]{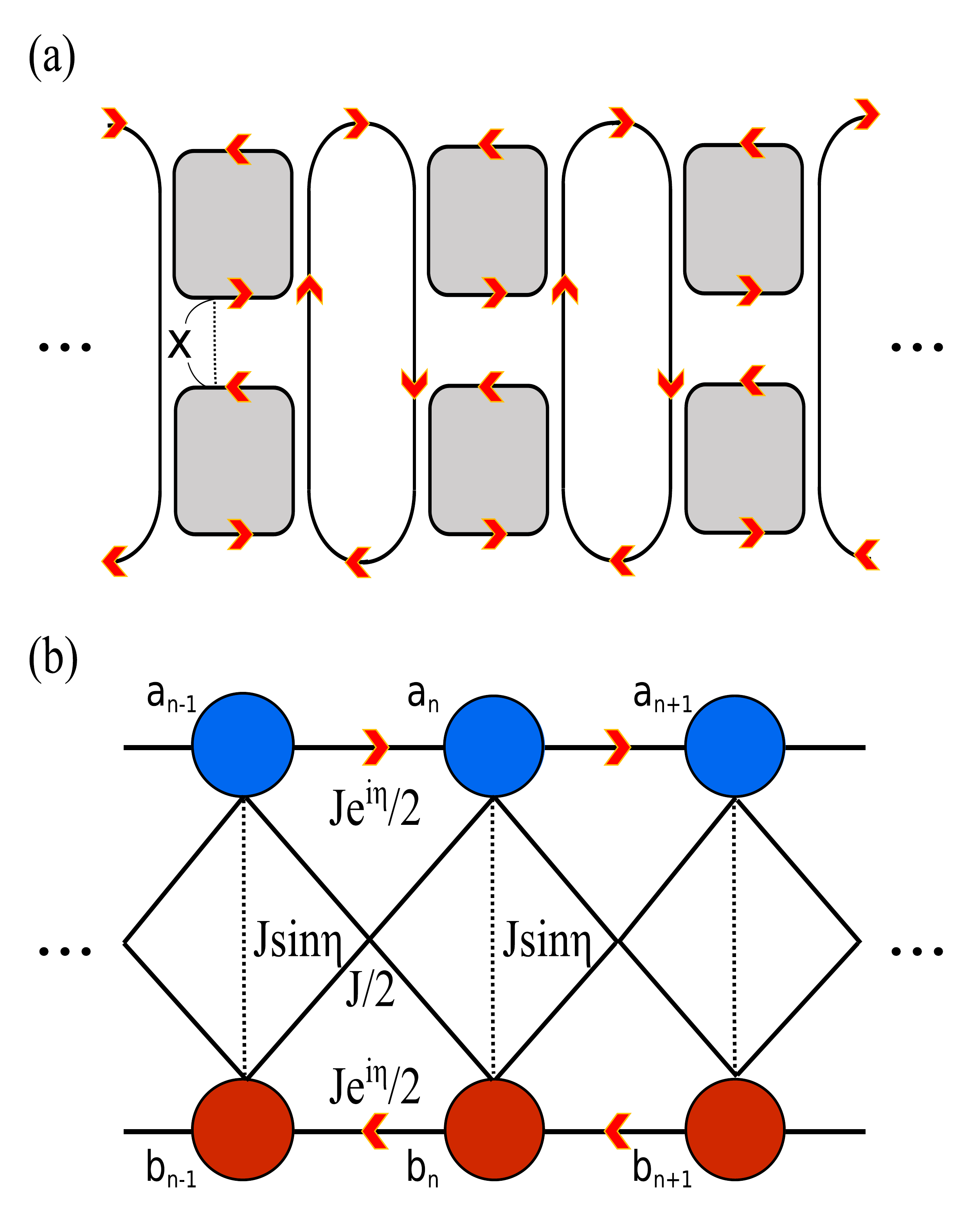}
\caption{(a) Schematic diagram of coupled resonator optical waveguide (CROW). Grey colour represents the cavity of each sublattice a and b. There is an off-resonant link between each site. (b) Schematic of tight binding model Eq.~\eqref{Hamiltonian}. Due to synthetic magnetic flux, the link mediates asymmetric coupling between each site with hopping phase $\eta$. Coupling strength between different sublattices within the same unit cell is $J \sin \eta$, and $J/2$ between neighbouring cells. }
\label{Schematics}
\end{figure}
To introduce a spin-like degree of freedom, we will use the two-leg ladder illustrated in Fig.~\ref{Schematics}(a). The two sublattices formed by resonant site rings are analogous to spin up and spin down states. In contrast to the approach of Refs.~\cite{Hafezi11,Hafezi13} and similar to the scheme introduced in Ref.~\cite{Leykam18}, coupling is mediated via a single off-resonant link ring per unit cell, which provides strong next-nearest-neighbor coupling emulating the spin-orbit coupling required for helical transport. To tune the relative strength of the intra-leg (spin-preserving) and inter-leg (spin-flipping) couplings, we allow for a variable separation $x$ between the two sublattices, which controls a phase delay $\eta = 2 \pi n_{\mathrm{eff}} x / \lambda$ accumulated in the link rings. Further details of the scattering matrices describing coupling between site and link rings may be found in Appendix A. When the effective inter-ring coupling strength $J$ is much smaller than FSR, i.e. $\theta \coloneqq \sqrt{4\pi J/\mathrm{FSR}} \ll 1$, light propagation through such an array is well-approximated by the tight binding Hamiltonian~\cite{Hafezi11,Hafezi13,Leykam18}
\begin{equation}
\begin{aligned}
&\hat{H}_{0} = \sum_{n} \left(\hat{H}_{a} + \hat{H}_{b} + \hat{H}_{ab} + \hat{H}^{\dagger}_{ab}\right), \\&\hat{H}_{a} = \frac{J}{2}\hat{a}^{\dagger}_{n}\left( e^{-i\eta} \hat{a}_{n-1} + e^{i\eta} \hat{a}_{n+1} \right), \\&\hat{H}_{b} = \frac{J}{2}\hat{b}^{\dagger}_{n}\left( e^{i\eta} \hat{b}_{n-1} + e^{-i\eta} \hat{b}_{n+1} \right),
\\&\hat{H}_{ab} = J \hat{a}^{\dagger}_{n}\left( \sin\eta \ \hat{b}_{n} + \frac{1}{2}\left( \hat{b}_{n-1} + \hat{b}_{n+1}\right)  \right),
\end{aligned}
	\label{Hamiltonian}       	
\end{equation}
illustrated in Fig.~\ref{Schematics}(b). Here, $\hat{a}^{\dagger}_n$ and $\hat{b}^{\dagger}_n$ are creation operators for the upper and lower legs, the integer $n$ indexes the lattice sites, and we measure energies (frequencies) with respect to a resonance frequency of the site rings. Note that for full generality we have used second quantization notation for $\hat{H}_0$, applicable to both classical and quantum states of light. In the following we will focus on the semi-classical limit, i.e. propagation of single photon or coherent states.

The eigenvalues $\omega$ of $\hat{H}_0$ correspond to resonant frequencies of the CROW. Since $\hat{H}_0$ is the Hamiltonian of a periodic lattice, its eigenstates are Bloch waves $\psi^{(j)}_{n}(k) = |u_{j}(k)\rangle e^{i k n}$ where $j$ is the band index. Fourier transforming Eq.~\eqref{Hamiltonian}, we obtain the Bloch Hamiltonian,
	\begin{equation}
	\begin{aligned}
		\hat{H}_{0} &= \sum_{k} (\hat{a}^{\dagger}_{k}, \hat{b}^{\dagger}_{k}) \hat{\mathcal{H}}_{0}(k) \left( \begin{array}{ccc}
		\hat{a}_{k} \\
		\hat{b}_{k}
		\end{array} \right) \\&= J\sum_{k} (\hat{a}^{\dagger}_{k},  \hat{b}^{\dagger}_{k})\left( \begin{array}{ccc}
		\cos(k+\eta) & \sin\eta + \cos k \\
		\sin\eta + \cos k & \cos(k-\eta)
		\end{array} \right)\left( \begin{array}{ccc}
		\hat{a}_{k} \\
		\hat{b}_{k}
		\end{array} \right),
	\end{aligned}
		\label{Hamiltonian matrix}
	\end{equation}
where $\hat{a}_{k} \coloneqq \sum_{n}\hat{a}_{n} e^{i k n}/\sqrt{N}$, $\hat{b}_{k} \coloneqq \sum_{n}\hat{b}_{n} e^{i k n}/\sqrt{N}$ for the given system size $N$ and $k \in [-\pi,\pi]$ is the crystal momentum. As a two band (level) system, $\hat{\mathcal{H}}_0(k)$ is isomorphic to the Hamiltonian of a spin 1/2 particle and can be written compactly as $\hat{\mathcal{H}}_0 = J\textbf{d}\cdot\boldsymbol{\hat{\sigma}}$, where $\boldsymbol{\hat{\sigma}} = (\hat{I}_{2},\hat{\sigma}_{x},\hat{\sigma}_{y},\hat{\sigma}_{z})$, $\hat{I}_{2}$ is 2 by 2 identity matrix, and $\sigma_{j}$ $(j=x,y,z)$ are Pauli matrices which compose of $\mathfrak{su}(2)$ algebra associated with the vector \textbf{d}, $\textbf{d} = (d_{0},d_{x},d_{y},d_{z})$. The nonzero components of $\textbf{d}$ are
\begin{equation}
\begin{aligned}
    &d_{0} = \cos\eta\cos k, \\&d_{x} = \sin\eta + \cos k, \\& d_{z} = -\sin\eta\sin k,
\end{aligned}
\label{Bloch vector}
\end{equation}
yielding the two band single particle spectrum 
\begin{equation}
\omega_{\pm} (k) = d_0 \pm \sqrt{d_x^2 + d_z^2}.
\end{equation}
The corresponding eigenstates of $\hat{\mathcal{H}}$ are
\begin{equation}
    \begin{cases}
    |u_{+}(k)\rangle = \cos\left(\frac{\theta}{2}\right) |a\rangle + \sin\left(\frac{\theta}{2}\right) |b\rangle, \\  |u_{-}(k)\rangle = \sin\left(\frac{\theta}{2}\right) |a\rangle - \cos\left(\frac{\theta}{2}\right) |b\rangle,
    \end{cases}
    \label{eigenstate}
\end{equation}
where $\theta = \cos^{-1}\left(d_{z}/\sqrt{d^{2}_{x}+d^{2}_{z}}\right)$, $|a\rangle = (1,0)^{T}$ and $|b\rangle = (0,1)^{T}$ indicate pseudospin states corresponding to sublattice degree of freedom. 
The meaning of each term in Eq.~\eqref{Bloch vector} is as follows: $\hat{I}_{2}d_{0} $ describes the symmetric part of the intra-leg coupling, determining the bare effective mass $(\partial^{2}d_0 / \partial k^{2})^{-1}$. $\hat{\sigma}_x \sin \eta $ is analogous to a Zeeman magnetic field applied parallel to the ladder. $\hat{\sigma}_x \cos k $ and $d_z \hat{\sigma}_z$ describe intrinsic and Rashba-like spin-orbit couplings respectively. Crucially, the relative strengths of these three terms are tunable via the phase delay $\eta$, which allows the realization of a few interesting tight binding models.
\begin{figure}
\centering
\includegraphics[width=\linewidth, height=6cm]{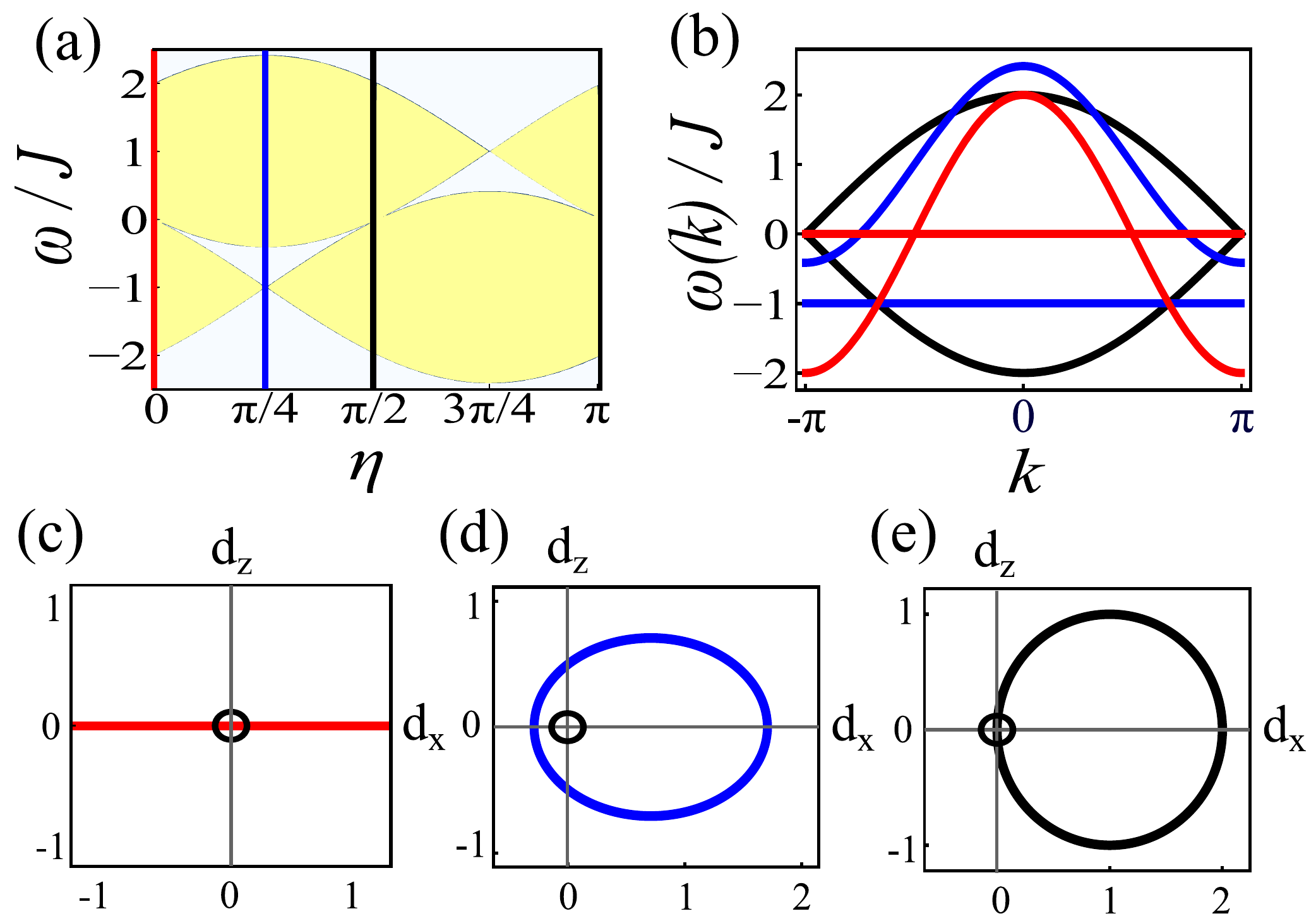}
\caption{(a) Bulk bands (shaded regions) as a function of the phase delay $\eta$, obtained from the tight binding model. Vertical lines indicate $\eta$ values of interest:  $\eta = 0$ (red), $\eta = \pi/4$ (blue) and $\eta = \pi/2$ (black). (b) Band dispersion diagrams of three specific $\eta$ values from (a). (c-e) Parameter plots of vector \textbf{d} in $(x,z)$-plane with singular (gap closing) point at $(d_{x},d_{z}) = 0$: (c) $\eta = 0$; (d) $\eta = \pi/4$; (e) $\eta = \pi/2$.}
\label{Band diagram}
\end{figure}

Fig.~\ref{Band diagram}(a) plots the spectrum of $\hat{H}_0$ as a function of the phase delay parameter $\eta$. Increasing $\eta$ from zero initially opens a gap in the spectrum, which reaches a maximum size at $\eta = \pi/4$ before vanishing again at the critical point $\eta = \pi/2$. The dispersion for the three limits of interest shown in Fig.~\ref{Band diagram}(b): First, when $\eta = 0$ the Zeeman and Rashba terms vanish and the model reduces to the cross-stitch lattice model introduced in Ref.~\cite{flach14}. It has a zero energy flat band embedded in a dispersive band $\omega_{+}(k) = 2J\cos k$. Second, when $\eta = \pi/4$ or $3\pi/4$, we obtain a sawtooth lattice-like band structure~\cite{Zhang15} with flat ($\omega_{-}(k) = \mp J$) and dispersive ($\omega_{+}(k) = \pm J(1+\sqrt{2}\cos k)$) bands separated by a gap.  Finally, when $\eta = \pi/2$, there is a band-crossing at $k= \pm\pi$ and the ladder has the simple dispersion relation $\omega_{\pm}(k) = \pm2J\cos(k/2)$, with $k= \pm\pi$ forming a critical point at which the amplitude of the vector \textbf{d} vanishes.

Except for the critical values $\eta = 0, \pi/2$, the eigenvectors of $\hat{\mathcal{H}}_0(k)$ have a nontrivial winding in $k$ due to the competition between the intrinsic and Rashba-like spin-orbit coupling terms. In particular, the Bloch Hamiltonian has the  symmetry $d_y = 0$, requiring its Bloch wave eigenstates to be confined to the $(\hat{\sigma}_x,\hat{\sigma}_z)$ plane of the Bloch sphere, as seen in Eq.~\eqref{eigenstate}. Figs.~\ref{Band diagram}(c-e) plot the components $(d_{x},d_{z})$ with the polar angle corresponding to the angle $\theta$ of the eigenstates for the three cases. When $\eta = 0$, the Rashba-like spin-orbit coupling vanishes and the eigenvectors become $k$-independent: the flat band modes are antisymmetric, $\psi^{(F)}_{n} = \frac{1}{\sqrt{2}}(1,-1)^{T} e^{i k n}$, while the dispersive band modes are symmetric, $\psi^{(D)}_n = \frac{1}{\sqrt{2}}(1,1)^{T} e^{i k n}$. For $0< \eta < \pi/2$, the eigenstates encircle the origin once. At the critical point $\eta = \pi/2$ the trajectory remains circular, but the circle touches the origin at $k = \pi$, corresponding to the gap closing. 

In these circular trajectories the $\hat{\sigma}_z$ spin axis is special, because from the form of $d_z$ we see that reversal of the momentum $k \rightarrow -k$ necessarily flips the $z$ component of the spin, in contrast to the $x$ component which is an even function of $k$. Therefore, when $d_x = 0$ we obtain spin-momentum locked eigenstates. This condition is satisfied when $\cos k = - \sin \eta$, or equivalently at energies $\omega = -J\sin 2\eta,\ 0$. Note that in the vicinity of $\eta = \pi/4$ or $3\pi/4$, spin-momentum locking disappears in the flat band since all wavevectors become degenerate.

So far we have assumed a perfectly periodic lattice Hamiltonian $\hat{H}_{0}$. In practice, however, fabrication imperfections are inevitable and we need to take sources of disorder into account: (1) Sidewall roughness of the resonators lowers their quality factors by introducing scattering losses $\kappa \sim 2$ GHz. (2) Misalignment of the resonator positions will lead to disorder in the inter-resonator coupling strengths, $\Delta J \sim 1$ GHz. (3) Most significantly, misalignment of the resonance frequencies leads to on-site disorder $\Delta \omega \sim 30$ GHz~\cite{Canciamilla10, Hafezi13, Mittal14, Mittal18}. It is not negligible compared to the hopping strength $J \sim 20$ GHz~\cite{Canciamilla10}. For simplicity, we will focus on the dominant latter term, which is described by the disorder Hamiltonian,
\begin{equation}
\begin{aligned}
\hat{V}_{\epsilon} &= \sum_n \left( V^{(a)}_{n,\epsilon} \hat{a}_n^{\dagger} \hat{a}_n + V^{(b)}_{n,\epsilon} \hat{b}_n^{\dagger} \hat{b}_n \right), \label{eq:disorder}
\end{aligned}
\end{equation}
here $\epsilon$ indexes different disorder realizations. We will assume that the disorder is statistically homogeneous, with site detunings $V_{n,\epsilon}^{(r)}$ $(r = a$ or $b$) uniformly distributed in the interval $[-\frac{W}{2},\frac{W}{2}]$, where $W$ is the disorder strength. Formally, for the given probability distribution about each disorder realization $p_{\epsilon}$, $\overline{\hat{V}} \coloneqq \int d\epsilon p_{\epsilon} \hat{V}_{\epsilon} = 0$.  For generality, we will allow for local correlations leading to different disorder symmetries, 
\begin{equation}
\begin{aligned}
&\int d\epsilon \ p_{\epsilon} V^{(r)}_{m,\epsilon} V^{(s)}_{n,\epsilon} \coloneqq \overline{V^{(r)}_{m,\epsilon} V^{(s)}_{n,\epsilon}} \\&= \frac{W^{2}}{12}\delta_{mn} \times
\begin{cases}
\delta_{rs} \ \text{(asymmetric)}, \\ 1 \ \text{(symmetric)},\\-1+2 \delta_{rs} \ \text{(anti-symmetric)},
\end{cases}
\end{aligned}
\label{correlation}
\end{equation}
where $(m,n)$ are site indices, $(r,s)$ index the sublattices $(r,s = a$ or $b)$, and $\delta_{rs}$ is the Kronecker delta function.  

\section{Anderson localization length}

The phenomenon that the wave is localized under the static random disorder with typical length scale $\xi$ is Anderson localization. In tight binding Hamiltonian such as Eq.~\eqref{Hamiltonian matrix}, the static disorder Eq.~\eqref{eq:disorder} generically leads to Anderson localization~\cite{Kramer93}. Thus, the Anderson localization length $\xi$ sets an upper bound on the length of the CROW; beyond this distance no appreciable transmission is possible, even in the absence of scattering losses. We will show analytically and numerically that spin-momentum locking leads to a strong enhancement of the Anderson localization length compared to a conventional CROW with the same group velocity $v_g$, enabling buffering of signals for a longer time.

In one-dimensional systems, the Anderson localization length $\xi$ is related to the scattering time $\tau$ as $\xi = 2 v_g \tau$~\cite{Kramer93}. In the presence of weak disorder, we can use the Born approximation to analytically calculate $\tau$ and hence $\xi$. Namely, in the presence of weak disorder $\hat{V}$ the Green's function of the system $\hat{G}$ can be expanded using perturbation theory as: 
\begin{equation}
\hat{G} = \hat{G}_{0} + \hat{G}_{0}\hat{V}_{\epsilon}\hat{G}_{0} + \hat{G}_{0}\hat{V}_{\epsilon}\hat{G}_{0}\hat{V}_{\epsilon}\hat{G}_{0} + ...
\label{interacting green fucntion}
\end{equation}
where $\hat{G}_{0} = \frac{1}{E - \hat{H}_{0} + i0}$ is the Green's function in the absence of disorder. Under the Born approximation the self energy $\Sigma$ defines the energy shift of the plane wave eigenstates due to the disorder~\cite{Lancaster14}, $\hat{\Sigma}_{\epsilon}(k,E) \coloneqq \sum_{k'} \hat{V}_{\epsilon}(-k,k')\hat{G}_{0}(k',E)\hat{V}_{\epsilon}(k',k)$. The plane wave eigenstates acquire a finite lifetime, the elastic scattering time $\tau$, where $(\tau_{j}(k))^{-1} = -\text{Im}\overline{\langle \hat{\Sigma}(k)\rangle}_{j}/\pi,\ \langle ... \rangle_{j} = \langle u_{j}(k) | ... |u_{j}(k) \rangle$ is the projection onto the Bloch state ($j = +$ or $-$), and $\overline{\hat{\Sigma}} = \int d\epsilon p_{\epsilon} \hat{\Sigma}_{\epsilon}$ is the disorder-averaged self-energy~\cite{Akkermans07}. Since $\overline{\hat{V}} = 0$, this self-energy term arises at second order in $\hat{V}_{\epsilon}$. Now, let us obtain the scattering time for the first band. The diagonal component which is projected onto one specific Bloch state e.g. $|u_{+}(k)\rangle e^{i k n}$ of averaged interacting Green's function is thus 
\begin{equation}
\begin{aligned}
&\overline{\langle u_{+}(k)| \hat{G}(k,E) |u_{+}(k) \rangle} \approx  \frac{1}{E - \omega_{+}(k) + i0} \\& + \left(\frac{1}{E - \omega_{+}(k) + i0}\right)^{2} \overline{\langle u_{+}(k)|\hat{\Sigma}(k,E)|u_{+}(k) \rangle}.
\end{aligned}
\label{green function}
\end{equation}
Where we apply Born approximation up to order of $V^{2}$. From the Sokhotski$-$Plemelj theorem \cite{weinberg95},
\begin{equation}
 	\frac{1}{E-\omega(k)+i0} = P\frac{1}{E-\omega(k)} - i\pi \delta(E-\omega(k)),
 \label{SP formula}
 \end{equation}
where P is a Cauchy principle value which is real. Since its imaginary part describes scattering effect by disorder, let us take a closer look at the imaginary part by substituting $E$ into $\omega(k)$.
Using the identity of delta function, one can obtain the relation between scattering time of one specific band. Namely, using the group velocity for $\omega(k) = \omega_{+}(k)$, 
\begin{equation}
 	\frac{1}{\tau_{+}(k)} =  \left|\frac{d\omega (k)}{dk}\right|^{-1} \left(\int dk' \overline{|\langle u_{+}(k)|\hat{V}|u_{+}(k')\rangle|^{2}}\delta(k + k')\right). 
 	\label{scattering time}
\end{equation}
For the most important case of asymmetric disorder, we calculate the Anderson localization length for phase delays $\eta = \pi/4$ and $\eta = \pi/2$. The calculation, detailed in Appendix B, yields
\begin{equation}
	\frac{\xi(\omega)}{24} = \begin{cases} 
     \frac{(\omega+J)^{2}(2J^{2}-(\omega-J)^{2})}{W^{2}\omega^{2}}  \small{\begin{aligned}
     (&\eta = \pi/4, \\&(1-\sqrt{2})J \leq \omega \leq (1+\sqrt{2})J),    
     \end{aligned}} \\\\ \frac{2J^{2}(4J^{2}-\omega^{2})}{W^{2}\omega^{2}} \quad \small{(\eta = \pi/2, -2J \leq \omega \leq 2J)}.
     \label{localization length}
    \end{cases}
\end{equation}
\begin{figure}
\centering
\includegraphics[width=\linewidth, height=7cm]{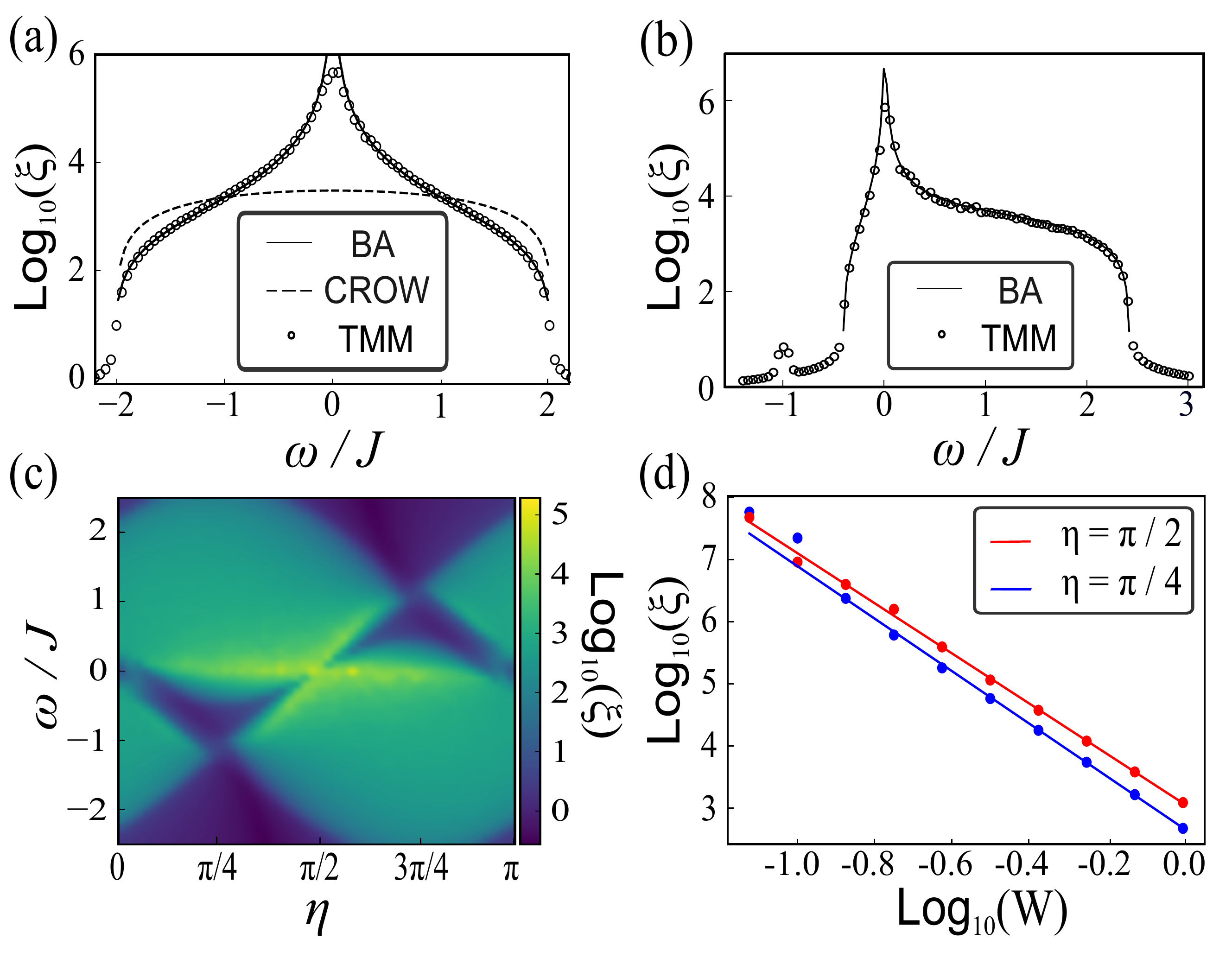}
\caption{(a) Localization length $\xi$ for asymmetric disorder when $\eta = \pi/2$. Analytic result (Born approximation (BA)) is represented by solid line and numeric (Transfer matrix method (TMM)) as dots. Comparison with simple CROW using Born approximation (CROW) is included as a dashed line. (b) Localization length $\xi$ for asymmetric disorder when $\eta = \pi/4$. (c) Localization length as a function of the phase delay parameter $\eta$. For (a) and (b) we use $W = 0.25J$, and we use $W = 0.5J$ for (c). (d) Power law scaling of the localization length for weak disorder. When $\eta = \pi/2$ (red), the power exponent $\nu \approx -4.1$. For $\eta = \pi/4$ (blue), $\nu \approx -4.03$.}
\label{localization plot1}
\end{figure}
\begin{figure}
\centering
\includegraphics[width=\linewidth, height=7cm]{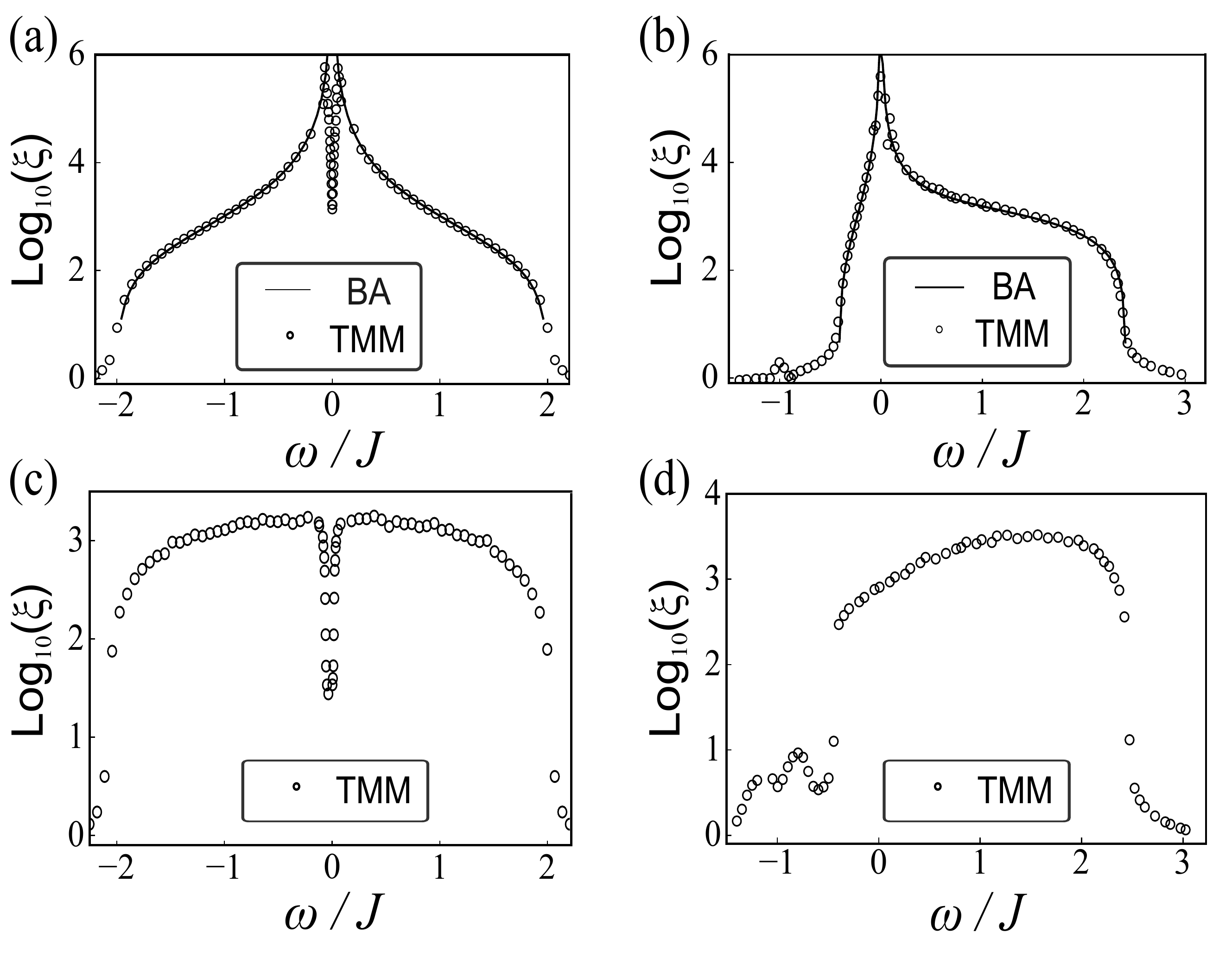}
\caption{(a-d) Localization length $\xi$ with respect to the detuning frequency $\omega$ in the presence of two types of disorder: (a) Symmetric, $\eta = \pi/2$ and $W = 0.25J$ (b) Symmetric, $\eta = \pi/4$ and $W = 0.25J$ (c) Anti-symmetric, $\eta = \pi/2$ and $W = 1.0J$ (d) Anti-symmetric, $\eta = \pi/4$ and $W = 1.0J$.}
\label{localization plot2}
\end{figure}

When $\eta = \pi/4$ or $\eta = \pi/2$, a divergence occurs at $\omega = 0$. This is the anticipated disorder robustness due to spin-momentum locking. Namely, at this point both the disorder $\hat{V}$ and Bloch Hamiltonian $\hat{H}$ are diagonal in the sublattice basis, i.e. the disorder cannot flip the spin and backscattering vanishes in the Born approximation. 
Meanwhile, at the band edges $\omega = (1 \pm \sqrt{2})J$ of $\eta = \pi/4$ and at $\omega = \pm 2J$ of $\eta = \pi/2$, the minimum localization length is obtained, because the group velocity vanishes. Note that for the flat band case, we cannot obtain localization length using Born approximation due to zero group velocity. 

We also calculate $\xi$ numerically using the transfer matrix method (details given in Appendix C), comparing against the analytical results in Figs.~\ref{localization plot1}(a,b). We obtain excellent agreement between the two, except in the vicinity of $\omega = 0$, where $\xi$ is large but remains finite; higher order terms smooth out the divergence appearing under the Born approximation. For comparison, the localization length of a conventional 1D CROW without spin-momentum locking is $\xi(\omega) = 48(4J^{2}-\omega^{2})/W^{2}$ under the Born approximation~\cite{Kramer93}, plotted as a dashed line in Fig.~\ref{localization plot1}(a). One can observe a substantial enhancement of the localization length for $|\omega| < J$, despite both systems sharing the same dispersion relation. Fig.~\ref{localization plot1}(b) similarly shows the maximum localization length at $\omega = 0$ as we expected. Unlike $\eta = \pi/2$, it is asymmetric with respect to zero detuning due to asymmetric band dispersion.

Fig.~\ref{localization plot1}(c) shows the numerically-obtained localization length for other phase delays $\eta$, revealing a strong enhancement of $\xi$ whenever we have the spin-momentum locking of the Bloch waves, i.e. at the energies $\omega = -J\sin 2\eta$, $0$ identified in the previous section (yellow regions), not just in the special cases $\eta = \pi/4, \pi/2$. Thus, this strong enhancement of $\xi$ is robust to detunings in the phase delay $\eta$.

Because the first order Born approximation gives a divergent localization length, the finite $\xi$ observed numerically must be due to higher order terms. We compute the scaling of $\xi$ with the disorder strength $W$ at zero detuning in Fig.~\ref{localization plot1}(d), obtaining a non-trivial power law $\xi \propto W^{-4}$. One can guess that this power factor originates from the second order Born approximation, since it is equivalent to the square of the standard $\xi \propto W^{-2}$ law of the first order Born approximation. Since our system involves only two bands, it may be possible to obtain this power law analytically by solving a Fokker-Planck equation~\cite{Ihor17}. 

To better understand the origin of this enhancement of $\xi$ at zero detuning, we also consider the effect of different disorder symmetries. Under the Born approximation, we find that $\xi$ is halved for symmetric disorder, and diverges for anti-symmetric disorder (see Appendix B). Fig.~\ref{localization plot2} shows the corresponding numerical results. We obtain good agreement for symmetric disorder, while for anti-symmetric disorder $\xi$ is strongly enhanced but remains finite. Interestingly, in both cases when $\eta = \pi/2$ there is an anomalous dip in $\xi$ at zero detuning, indicative of nontrivial behaviour at higher orders due to multiple scattering (Figs.~\ref{localization plot2}(a,c)). Meanwhile, $\eta = \pi/4$ does not show the dip about zero detuning (Figs.~\ref{localization plot2}(b,d)). We expect it is due to the asymmetry in the dispersion.


\section{Pulse Propagation}

In the previous section we calculated the energy-dependent Anderson localization length, which describes the system under excitation by monochromatic (continuous wave) beams. This does not take into account how the coherence between different frequency components making up an optical pulse may or may not be preserved during propagation through the lattice. Namely, we integrate over scattering states assuming independent mode contributions. This process is incoherent since we do not take the relative phase of different state into account. What we obtained actually was the transition rate within the same band, which is time independent. However, this information is insufficient to describe the temporal evolution of wavepackets, which is generally affected by disorder-induced non-local correlation effects; moreover we cannot obtain phase information, which in turn gives the coherence between fields in the two sublattices.

In order to study the disorder impact on pulses with finite band width, as they occur in actual experiments and devices, we now consider wavepacket propagation in the time domain. This will allow us to assess the disorder-induced backscattering in two ways complementary to the previous analysis in terms of the localization length: first, by directly observing the appearance of backscattering peaks, and second, indirectly by tracking the purity evolution of the disorder-averaged state. In case of backscattering-free propagation, the purity decays to a characteristic plateau value, indicating the unavoidable disorder-induced dephasing \cite{Gneiting17}. Strong deviations from this plateau value, i.e., increasing overshooting, can then be taken as a signature of backscattering, since the latter also adds to the mixing of the disorder-averaged state and thus to its purity decay. In addition, such purity test allows us to assess the coherence properties of the disorder-averaged state.

First, let us complement our numerical investigation by deriving an effective temporal evolution equation for the disorder-averaged field state, which is valid in the limit of weak disorder \cite{Gneiting17A, Gneiting18disorder}. We begin with a recap of the general line of argument. Starting point are the (temporally evolving) field states $|\psi_{\epsilon}\rangle$ of individual disorder realizations. Equivalently, we can consider the corresponding density matrices $\rho_{\epsilon} = |\psi_{\epsilon}\rangle \langle \psi_{\epsilon}|$, which underlie the definition of the disorder-averaged state $\overline{\rho} = \int d\epsilon\ p_\epsilon \rho_{\epsilon}$. These density matrices obey Liouville equation~\cite{Breuer02},
\begin{equation}
    i\partial_{t}\rho_{\epsilon}(t) = [\hat{H}_{\epsilon},\rho_{\epsilon}(t)],
    \label{Liouville equation}
\end{equation}
where $\hat{H}_{\epsilon} = \hat{H}_{0} + \hat{V}_{\epsilon}$, and $[\hat{H}_{\epsilon},\rho_{\epsilon}(t)] \coloneqq  \hat{H}_{\epsilon}\rho_{\epsilon}(t) - \rho_{\epsilon}(t)\hat{H}_{\epsilon}$. To proceed towards a master equation for the disorder-averaged state, we now separate the state $\rho_{\epsilon}(t)$ into two parts: 1) the ensemble-averaged state $\bar{\rho}(t)$ and 2) a disorder-induced fluctuation $\Delta \rho_{\epsilon}(t)$.
 From Eq.~\eqref{Liouville equation}, one can then derive coupled evolution equations for the average part and the individual offsets,
 \begin{align}
i\partial_{t}\bar{\rho}(t) =& [\hat{H}_{0},\bar{\rho}(t)]+\int d\epsilon \ p_{\epsilon} [\hat{V}_{\epsilon},\Delta \rho_{\epsilon}(t)], \nonumber \\
i\partial_{t}\Delta\rho_{\epsilon}(t) =& [\hat{H}_{\epsilon},\Delta\rho_{\epsilon}(t)]+[\hat{V}_{\epsilon},\bar{\rho}(t)] \nonumber \\
&-\int d\lambda\  p_{\lambda} [\hat{V}_{\lambda},\Delta \rho_{\lambda}(t)].
\label{Liouville equation separate}
\end{align}
Solving the second equation of \eqref{Liouville equation separate}, and taking Born approximation up to $O(\hat{V}^{2})$ terms, we obtain a closed evolution for the average state~\cite{Gneiting17A},
\begin{equation}
\begin{aligned}
   i\partial_{t}\bar{\rho}(t) = [\hat{H}_{0},\bar{\rho}(t)]  -i \int_{0}^{t} dt' \int d\epsilon \ p_{\epsilon} [\hat{V}_{\epsilon},[\hat{\tilde{V}}_{\epsilon}(t'),\bar{\rho}(t)]],
\end{aligned}
\label{master equation}
\end{equation}
where $\hat{\tilde{V}}_{\epsilon}(t') \coloneqq \hat{U}_{t'}\hat{V}_{\epsilon}\hat{U}_{t'}^{\dagger}$. We remark that Eq.~\eqref{master equation} can be manifestly formulated in Lindblad form~\cite{Gneiting17A,Gneiting18disorder}. Moreover, we stress that the time integral indicates the non-Markovian, i.e., time-nonlocal nature of the disorder impact, and thus cannot be simplified, e.g., by taking the limit $t \rightarrow \infty$, without losing essential aspects of the disorder-induced evolution.

We now evaluate the general evolution equation~\eqref{master equation} for our specific system. As we are interested in the behaviour about spin-momentum locked points, it is most efficient  
to take the long wavelength-limit, in particular since the ensemble average still possesses translation symmetry. Concretely, we take the continuum limit, $\sum_{n} \rightarrow \int dx$. Then, the disorder operator, expressed in terms of the momentum basis, reads
 	\begin{equation}
 	\begin{aligned}
 	     	\hat{V}_{\epsilon} \coloneqq \int_{-\infty}^{\infty} \int_{-\infty}^{\infty} \int_{-\infty}^{\infty} dp dq dx \ e^{i(q-p)x} \ \hat{V}_{\epsilon}(x)  \otimes |p\rangle\langle q|,
 	\end{aligned}
 	\label{disorder potential continuous}
 	\end{equation}
 	where the representation of $\hat{V}_{\epsilon}$ in terms of sublattice basis $\lbrace |a\rangle,|b\rangle \rbrace$ is 
 	\begin{equation}
 	\hat{V}_{\epsilon}(x) = \left( \begin{array}{ccc}
 	V^{(a)}_{\epsilon} & 0 \\
 	0 & V^{(b)}_{\epsilon}
 	\end{array} \right).
 	\end{equation}
 To take the continuum limit, we need to introduce a characteristic length scale for the disorder, via the spatial correlation function $C_{ab}(x-x') \coloneqq \int_{\epsilon}\ p_{\epsilon} V^{(a)}_{\epsilon}(x)V^{(b)}_{\epsilon}(x')$ associated with its Fourier transformation $C_{ab}(x-x') \coloneqq \int_{\infty}^{\infty} dq \ G_{ab}(q) \exp(iq(x-x'))$.
 The master equation~\eqref{master equation}, evaluated for asymmetric disorder, then reads
\begin{equation}
\begin{aligned}
 	&i\partial_{t}\bar{\rho}(t) = [\hat{H}_{0},\bar{\rho}(t)] -i\sum_{\beta = a,b} \int_{0}^{t} dt'  \int_{-\infty}^{\infty} dq \ G(q) \\& \times [\hat{P}_{\beta}\otimes \hat{W}_{q}  ,[\exp(i\hat{H}_{0}t')(\hat{P}_{\beta}\otimes \hat{W}_{-q})\exp(-i\hat{H}_{0}t') ,\bar{\rho}(t)]],
 	\label{asymmetric disorder}
\end{aligned}
 \end{equation}
where $\hat{P}_{\beta} = |\beta\rangle\langle \beta |$ ($\beta = a,b$) are projection operators on the sublattices, corresponding to the pseudospin part of $\hat{V}_{q} \coloneqq \hat{P}_{\beta} \otimes \hat{W}_{q}$, and  $\hat{W}_{q}$ is a momentum kick operator of the form,
\begin{equation}
 \hat{W}_{q} \coloneqq \int dp |p\rangle\langle p+q| = \exp(iq\hat{x}).   
\end{equation}
In the presence of symmetric and anti-symmetric disorder, the sum over sublattice projectors in  Eq.~\eqref{asymmetric disorder} is replaced by  $\hat{I}_{2}$ and $\hat{\sigma}_{z}$, respectively. Note that, if $\hat{H}_{0}$ exhibits a nonzero vanishing $\hat{\sigma}_{x}$ component, $\hat{\tilde{V}}_{-q}(t') \coloneqq \exp(i\hat{H}_{0}t')(\hat{P}_{\beta}\otimes \hat{W}_{-q})\exp(-i\hat{H}_{0}t')$ comprises $\hat{\sigma}_{x}$ and $\hat{\sigma}_{y}$ components which flip the spin, a prerequisite for backscattering. 
 	\label{anti-symmetric disorder}

We now use this framework to discuss the disorder-induced dephasing in the vicinity of zero detuning from spin-momentum locking for given $\eta$. To this end, we expand the Hamiltonian $\hat{H}_{0}$ in Eq.~\eqref{Hamiltonian matrix} about the spin-momentum locking points $k_{0} = \pm \cos^{-1}(\sin \eta)$,
\begin{equation}
\begin{aligned}
 &\hat{H}_{0}(\hat{p}) \approx + J\left[\hat{\sigma}_{x} \otimes \left\lbrace \pm\cos \eta \ \hat{p} - \frac{1}{2}\sin \eta \ \hat{p}^{2}\right\rbrace \right. \\& \left. -\hat{I}_{2} \otimes \left\lbrace\frac{1}{2}\sin 2\eta \pm\cos^{2}\eta \ \hat{p} + \frac{1}{2}\sin 2\eta \ \hat{p}^{2}\right\rbrace  \right. \\& \left. - \hat{\sigma}_{z} \otimes \left\lbrace \pm\sin\eta \cos\eta + \sin^{2}\eta \  \hat{p} \pm \frac{1}{2}\cos\eta \ \hat{p}^{2} \right\rbrace \right],
 \label{Hamiltonian expansion eta}
\end{aligned}
\end{equation}
where $\hat{p} \coloneqq \hat{k} - k_{0}$ is the shifted momentum operator associated with $\hat{k} = \int_{-\infty}^{\infty} dk\ k  \hat{a}^{\dagger}_{k}\hat{a}_{k}$. Note that this equation exhibits two relative signs for $\cos \eta$ because it assumes both positive and negative variation with respect to the spin-momentum locking point relating to two solutions for $k_{0}$. As mentioned above, the expansion order in $\hat{p}$ about the (spin-flipping) $\hat{\sigma}_{x}$ determines the degree of robustness against disorder.  
In particular, if and only if $\eta = \pi/2$, the $\hat{\sigma}_{x}$ contribution linear in $\hat{p}$ vanishes, and the quadratic order term becomes leading. Thus, we can conclude that, in the case of asymmetric and symmetric disorder, $\eta = \pi/2$ is the most robust point with respect to momentum deviations.

Let us now consider the case of $\eta = \pi/2$ to investigate the dephasing behaviour in the presence of different sublattice correlations. Again in the vicinity of the spin-momentum locking point, which lies at $k_0 = \pi$, $\hat{H}_{0}$ is then, up to quadratic order, given by
\begin{equation}
\hat{H}_{0}(k_{0} = \pi) \approx J\hat{\sigma}_{z} \otimes \hat{p} + J\hat{\sigma}_{x}\otimes \frac{\hat{p}^{2}}{2}.
\label{eq: Expansion Hamiltonian}
\end{equation}
Assuming small deviation of momentum, we can neglect the second order in $\hat{p}$ of Eq.~\eqref{eq: Expansion Hamiltonian}. Then, the master equation~\eqref{master equation} reads \cite{Gneiting17}
\begin{equation}
\begin{aligned}
&i\partial_{t}\bar{\rho}(t) = J[\hat{\sigma}_{z}\otimes\hat{p},\bar{\rho}(t)] \\&- 2it\int_{-\infty}^{\infty} dq \ G(q) \ \text{sinc}(qJt) \left\lbrace \bar{\rho}(t) - \hat{V}_{q} \bar{\rho}(t) \hat{V}_{-q} \right\rbrace. 
\end{aligned}
\label{Linear master equation}
\end{equation}
This equation~\eqref{Linear master equation} can be solved exactly for any sort of disorder. As we can observe in Eq.~\eqref{Linear master equation}, in the considered approximation, there is absence of backscattering in the sublattice basis $\lbrace |a\rangle,|b\rangle \rbrace$, since there is no spin mixing contribution regardless of the disorder characteristics. Meanwhile, the averaged state still undergoes dephasing due to incoherent contribution from disorder correlation. Below, we will use this to assess the disorder robustness in terms of the purity decay. In order to explain both disorder-induced dephasing and backscattering, we would have to take the
full approximated Hamiltonian~\eqref{eq: Expansion Hamiltonian} up to quadratic order into account, which exhibits a $\sigma_{x}$ term.
 Again, one can conclude that appearance of a term proportional to $\hat{\sigma}_{x}$ causes backscattering for every disorder correlation. 
\begin{figure}
\centering
\includegraphics[width=\linewidth, height=8.5cm]{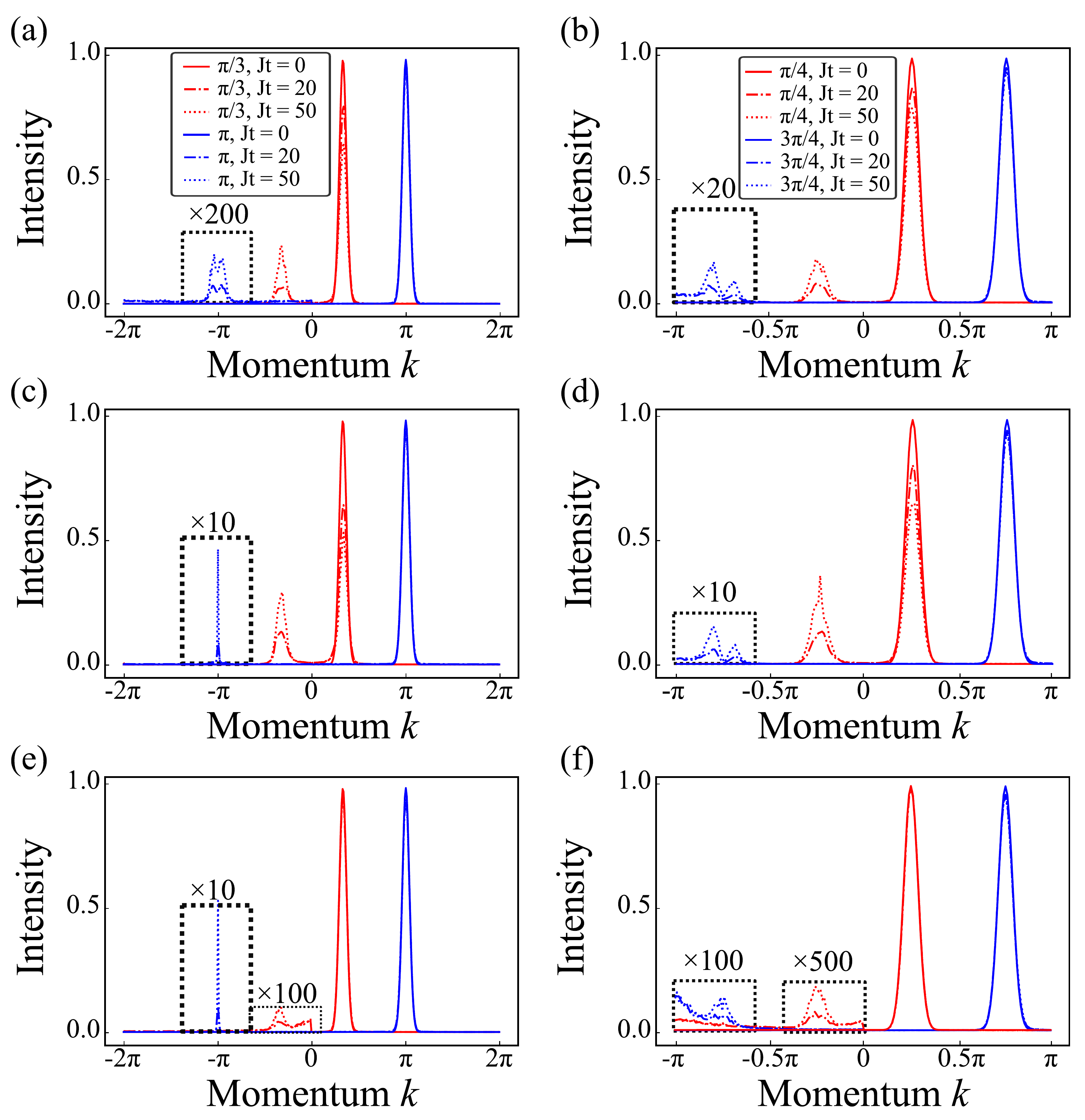}
\caption{Disorder-induced backscattering at normalized propagation times $Jt = 0, 20, 50$. We take the ensemble average over 100 realizations, moderate disorder $W = 0.5J$ and packet width $w = 6.0d$, where d is the lattice constant. Left column (a,c,e) shows the case $\eta = \pi/2$ for initial momentum $k_{0} = \pi$ (spin-momentum locking, blue) and $k_{0} = \pi/3$ (no spin-momentum locking, red). The right column (b,d,f) shows $\eta = \pi/4$ for $k_{0} = 3\pi/4$ (spin-momentum locking point, blue) and $k_{0} = \pi/4$ (no spin-momentum locking, red). First row: asymmetric disorder case. Second row: symmetric disorder. Third row: anti-symmetric disorder. Dashed regions are multiplied by indicated amounts to improve visibility.}
\label{fig:Wavepackets}
\end{figure}
Let us now turn to our numerical treatment in the time domain. We begin with the direct observation of backscattering peaks in the momentum distribution. To this end, we simulate propagation of wavepackets in the disordered tight binding model.  Fig.~\ref{fig:Wavepackets} shows their disorder-averaged momentum profiles for different propagation times and the given three types of disorder. Initial states are chosen Gaussian with width $w$, i.e., in momentum space, $\phi_{in}(k) = \sqrt{\frac{w^{2}}{2\pi}} \exp(-w^{2}(k-k_{0})^{2}/2)  $ and $|u_{+}(k)\rangle$ as the pseudospin part. We compare the evolved momentum profiles for the two initial momenta: at the spin-momentum locking point, and detuned from spin-momentum locking. We recover the suppression of backscattering for the asymmetric and symmetric disorder cases. At the same time, we can confirm that anti-symmetric disorder indeed gives rise, due to higher-order effects, to enhanced backscattering at the spin-momentum locking point. 

Let us look more closely at the effect of the choice of the parameter $\eta$. When $\eta = \pi/2$, the momentum distribution exhibits a symmetric backscattered wavepacket, since band dispersion is symmetric with respect to zero detuning, while choosing $\eta = \pi/4$ gives rise to an asymmetric profile about zero detuning, due to an asymmetric band profile. In the asymmetric disorder case, backscattering is more suppressed for $\eta = \pi/2$ than for $\eta = \pi/4$ (Figs.~\ref{fig:Wavepackets}(a,b)), but enhanced in the presence of symmetric disorders. In addition, with $\eta = \pi/2$ we encounter sharp peaks in the vicinity of the spin-momentum locking point for both the symmetric and the anti-symmetric disorder (Figs.~\ref{fig:Wavepackets}(c,e)), which can be traced back to the corresponding narrow dips in the respective localization length profiles. In contrast, the backscattering profiles remain smooth for $\eta = \pi/4$, in line with the respective localization length profiles (Figs.~\ref{fig:Wavepackets}(d,f)).

As an independent assessment of the disorder-induced backscattering of wavepackets, we now consider the purity evolution of the disorder-averaged state. The purity, which is defined as $\text{Tr}[\bar{\rho}^{2}]$ and measures the ``mixedness'' of a quantum state, indicates, when applied to the disorder-averaged state, to what extend states evolving under individual disorder realizations deviate from the unperturbed (disorder-free) evolving state \cite{Gneiting18disorder}. In particular, in the case of backscattering-free, dispersionless propagation, as approximated by our master equation~\eqref{Linear master equation}, it has been shown \cite{Gneiting17} that the purity evolves, due to unavoidable disorder-induced dephasing, into a characteristic plateau value given by
\begin{equation}
\begin{aligned}
 &\text{Tr}[\bar{\rho}^{2}](t) = 1 - (l^{2}C_{0}/ \pi v_{g}^{2})\\&(\sqrt{1+2(w/l)^{2}}[1-\exp(-(v_{g}t)^{2}/(l^{2}+2w^{2})]\\&-[1-\exp(-(v_{g}t)^{2}/l^{2}]\\&+\sqrt{\pi}(v_{g}t/l)(\text{erf}[v_{g}t/l]-\text{erf}[v_{g}t/\sqrt{l^{2}+2w^{2}}]),
\end{aligned}
\label{eq:analytic solution}
\end{equation}
where $l$ is the correlation length of Gaussian spatial correlation function, $C(x) = \frac{C_{0}}{2\pi}\exp (-(x/l)^{2})$, $v_{g}$ is the group velocity determined by the band dispersion, $C_{0} = W^{2}/12$ in our case, and $\text{erf}(x) \coloneqq 2/\sqrt{\pi} \int_{0}^{x} dt \exp(-t^{2})$ denotes the error function. Note that it does not depend on the correlation types of disorder for our spin-momentum locked initial state.

\begin{figure}
\centering
\includegraphics[width=\linewidth, height=8.5cm]{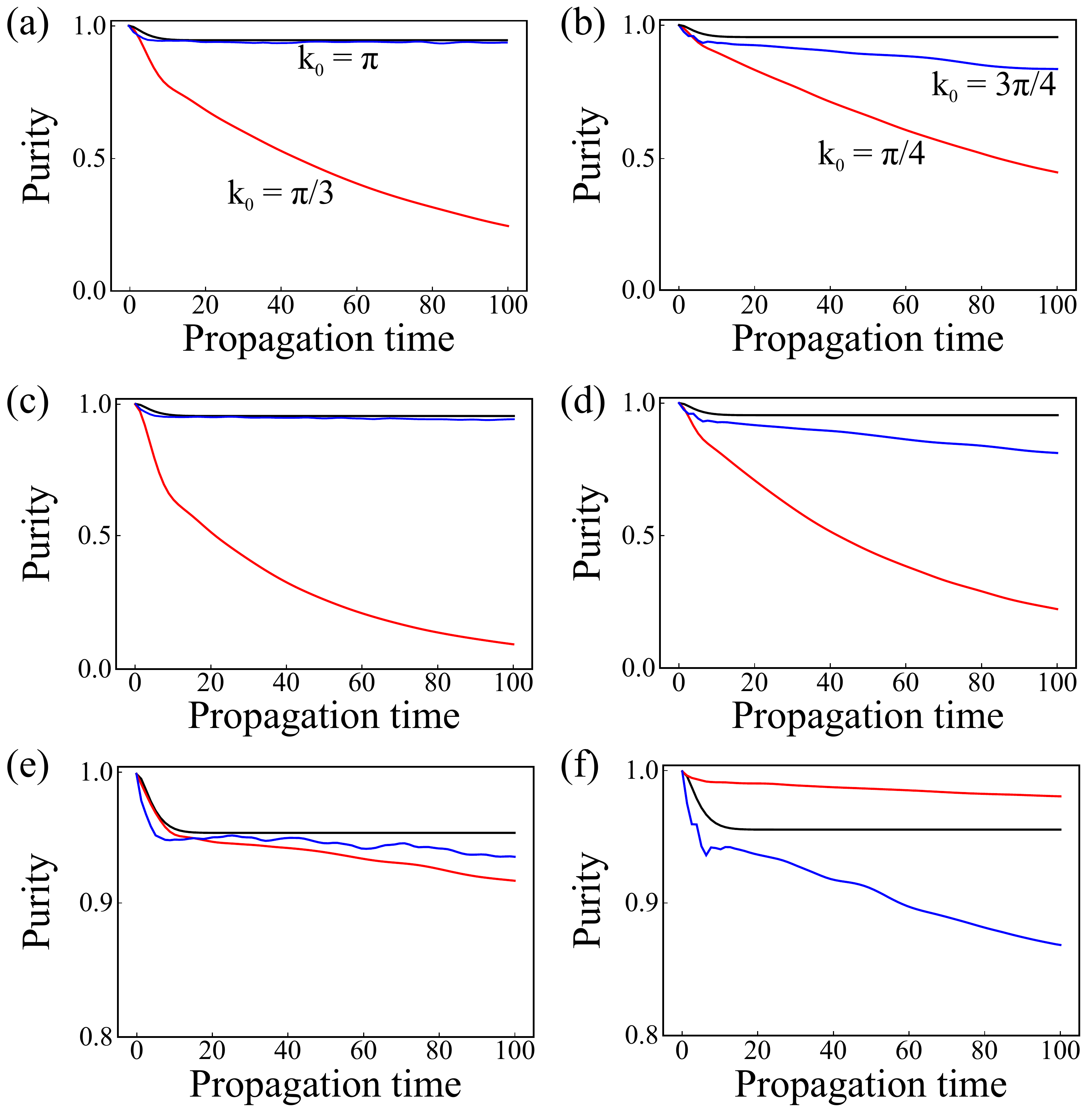}
\caption{Purity evolution for moderate disorder strength $W = 0.5J$ with ensemble size 400. Black line is the analytic result from~\cite{Gneiting17}. We use the effective correlation length $l = 2d$. Given phase delay is $\eta = \pi/2$ (left column: a,c,e), $\eta = \pi/4$ (right column: b,d,f). The red curves describe the field with initial momentum detuned from the spin-momentum locking point (left column: $k_{0} = \pi/3$ and right column: $k_{0} = \pi/4$). Blue curves evolve from initial momenta chosen about the spin-momentum locking point (left column: $k_{0} = \pi$ and right column: $k_{0} = 3\pi/4$). (a,b):  Asymmetric disorder, (c,d): Symmetric disorder, (e,f):  Anti-symmetric disorder. }
\label{fig:Plateau}
\end{figure}

Figure~\ref{fig:Plateau} plots the time evolution of the purity for the various cases. When $\eta = \pi/2$, the purity of the field at the spin-momentum locking point $k_{0} = \pi$ (blue lines in Figs.~\ref{fig:Plateau}(a,c,e)) indeed converges to the plateau value predicted by Eq.~\eqref{eq:analytic solution} (black line), confirming propagation with negligible backscattering. 
Here, we determined an effective correlation length $l = 2d$ ($d$: the lattice constant) by fitting (23) to the numerical solution. We find that in the detuned case (red line), the purity decay rapidly overshoots the predicted plateau value for backscattering-free transport, indicating that backscattering dominates the purity loss. In the correlated disorder cases, the plateau value is smaller compared to the asymmetric case (Figs.~\ref{fig:Plateau}(c,e)). Meanwhile, when $\eta = \pi/4$, the purity at spin-momentum locking point $k_{0} = 3\pi/4$ shows monotonic decay for all three cases (Figs.~\ref{fig:Plateau}(b,d,f)). Both the red and the blue curves decay beyond the prediction of backscattering-free transport ($k_{0} = 3\pi/4, \pi/4$), but the blue curve still shows slower decay than the red one in the presence of asymmetric and symmetric disorder, in agreement with the localization profiles in Fig.~\ref{localization plot1}(b) and Fig.~\ref{localization plot2}. Decay of purity can be understood from the localization profile in Figs.~\ref{localization plot1}(a,b). We show absolute value for each $\eta$ case is determined by the localization length, e.g. $\eta = \pi/2$, purity value indicates the largest value for the asymmetric disorder among three cases, and the smallest for the anti-symmetric correlated disorder. About $\eta = \pi/4$, since localization length for anti-symmetric correlated cases exhibits larger value than the symmetric correlated case, it shows larger value of purity also. For instance, $\eta = \pi/2$ exhibits a smooth profile with respect to the detuning parameter $\omega$, while the case $\eta = \pi/4$ shows a sharp peak at zero detuning. We can observe interesting behaviour for anti-symmetric disorder: The field without spin-momentum locking demonstrates more robust behaviour than the field with spin-momentum locking, cf. Figs.~\ref{fig:Plateau}(e,f). This is related to the localization lengths at the corresponding points, $k_{0} = \pi/3$ for $\eta = \pi/2$ and $k_{0} = \pi/4$ for $\eta = \pi/4$, being larger than for the spin-momentum locking points.

\section{Conclusion}
We have studied the design of disorder resistant helical transport in one-dimensional (1D) coupled resonator optical waveguides. We proposed a model which exhibits the spin-momentum locking of its one-dimensional bulk modes at critical energies and proved disorder resistance of this helical transport in two ways. Firstly, we have shown the enhancement of Anderson length compared to simple one-dimensional coupled ring resonator model. We computed the Anderson localization length analytically by calculating the self-energy using the Born approximation, obtaining excellent agreement with numerical results.  Second, we have studied the propagation of wavepackets in the time domain using a master equation formalism, showing that the spin-momentum locking minimizes backscattering and maximizes their purity. We have obtained the utmost disorder resistant behaviour occurs when $\eta = \pi/2$ via showing the existence of plateau for the value of purity with respect to propagation time. We believe this approach towards designing topological transport can be more efficient than conventional approaches based on higher dimensional lattices. Our approach can be applied to design disorder-resistant transport in quasi-1D optical waveguides.

\section*{Acknowledgement}

This research was supported by the Institute for Basic Science in Korea (IBS-R024-Y1).

\section*{Appendix}
\begin{appendix}
\section{Scatterting matrix formalism}

\indent The tight binding model Eq.~\eqref{Hamiltonian} approximates the more general scattering matrix description of the system in the limit of weak inter-resonator coupling. In this Appendix we will present the full scattering matrix model, similar to models previously employed in Refs.~\cite{Hafezi11,Leykam18}, and demonstrate that it gives similar results for the dispersion relation and localization length for typical experimental parameters.
\begin{figure}
    \centering
 \includegraphics[width=\columnwidth]{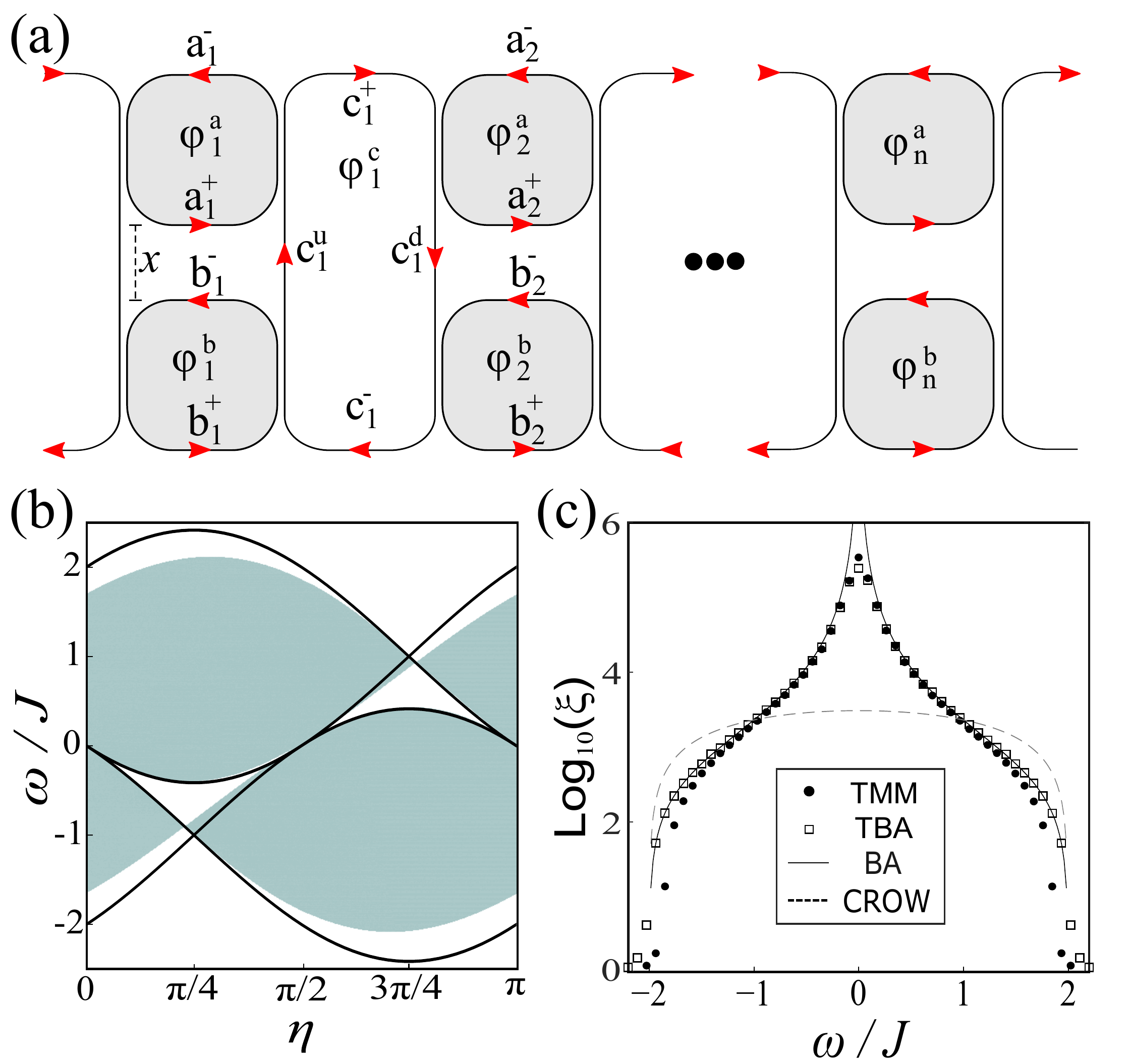}
    \caption{(a) Schematic diagram of CROW associated with optical field amplitude of each ring ($a^{\pm}_{n}, b^{\pm}_{n}$) and link ($c^{\pm}_{n}$), and accumulated phase ($\varphi^{a}_{n}$, $\varphi^{b}_{n}$). (b) Band structure of the array as a function of the coupling asymmetry $\eta$ obtained from the scattering matrices Eq.~\eqref{eq:s_matrices}. For comparison, the solid black lines denote the band edges obtained under the tight binding approximation. (c) Anderson localization length for $\eta = \pi/2$ (gapless limit) and asymmetric disorder $W=0.25J$. Tight binding  (TBA) and scattering matrices (TMM) give very similar results. Solid and dashed line indicate the analytic result under Born approximation (BA) and simple CROW model.}
    \label{fig:Smatrix_bstructure}
\end{figure}
Let $(a_n^{\pm},b_n^{\pm},c_n^{\pm})$ be optical field amplitudes in the ring segments as indicated in Fig.~\ref{fig:Smatrix_bstructure}(a), and $\varphi_n^j$ be round trip phases accumulated in each ring (ring-dependent to allow for disorder). $a_n^{\pm}$ and $b_n^{\pm}$ denote amplitudes in the resonant site rings, while $c_n^{\pm}$ are amplitudes in the anti-resonant link rings. The inter-ring couplings are parametrized by unitary scattering matrices $\hat{S} = \exp[-i \hat{\sigma}_x \theta ]$, where $\theta$ is the coupling angle, which relate the field amplitudes in neighboring rings as follows:
\begin{subequations}
\begin{align}
&\left( \begin{array}{c} a_n^- e^{-i \varphi^a_n/2} \\ c_{n}^{+} e^{-i (\varphi_n^c/2 - \eta )/2} \end{array} \right) =  \hat{S} \left( \begin{array}{c} a_n^+ e^{i \varphi^a_n/2}\\ c_{n}^u e^{i \eta/2}\end{array} \right), \\
& \left( \begin{array}{c} b_n^- e^{-i \varphi^b_n/2} \\ c_{n}^{u} e^{-i \eta/2} \end{array} \right) = \hat{S} \left( \begin{array}{c} b_n^+ e^{i \varphi^b_n/2} \\ c_{n}^- e^{i (\varphi_n^c/2 - \eta)/2}\end{array} \right), \\
&\left( \begin{array}{c} a_{n+1}^+ e^{-i \varphi^a_{n+1}/2} \\ c_{n}^{d} e^{-i \eta/2} \end{array} \right) =  \hat{S} \left( \begin{array}{c} a_{n+1}^- e^{i \varphi^a_{n+1}/2} \\ c_{n}^+ e^{i (\varphi_n^c/2 -\eta)/2} \end{array} \right),\\
&\left( \begin{array}{c} b_{n+1}^+ e^{-i \varphi^b_{n+1}/2} \\ c_{n}^{-} e^{-i (\varphi_n^c/2 -\eta)/2} \end{array} \right) = \hat{S} \left( \begin{array}{c} b_{n+1}^- e^{i \varphi^b_{n+1}/2} \\ c_{n}^d  e^{i \eta/2} \end{array} \right).
\end{align}
\label{eq:s_matrices}
\end{subequations}
Recall $\eta = 2\pi n_{\mathrm{eff}} x / \lambda$ is the coupling asymmetry induced by the site rings' offset, $n_{\mathrm{eff}}$ is the effective refractive index of the rings, and $\lambda$ is the free space wavelength. For a frequency detuning $\omega$ from resonance, the round trip phases are
\begin{subequations}
\begin{align} 
\varphi_n^{a,b} = 2\pi (\delta \varphi_n^{a,b} + \omega )/\mathrm{FSR},\\
\varphi_n^{c} = 2\pi ( 1/2 + \delta \varphi_n^c + 2 \omega )/\mathrm{FSR},
\end{align}
\end{subequations}
where $\mathrm{FSR}$ is the rings' free spectral range, $\delta \varphi_n^j$ describe the disorder in the ring resonant frequencies, and the $1/2+2\omega$ term in the second equation accounts for the longer length and anti-resonance of the link rings. Note that we include disorder in the link rings, $\delta\varphi^{c}_{n}$, which corresponds to (weak) coupling disorder in the tight binding model. The tight binding model Eq.~\eqref{Hamiltonian} can be obtained by solving Eq.~\eqref{eq:s_matrices} perturbatively in the weak coupling limit $\theta = \sqrt{4\pi J/\mathrm{FSR}} \ll 1$, similar to Ref.~\cite{Leykam18}

To compare the predictions of the scattering matrix and tight binding models, Eq.~\eqref{eq:s_matrices} can be rearranged into a transfer matrix that propagates a field at fixed frequency $\omega$ from unit cell $n$ to cell $n+1$. With this transfer matrix we compute the Bloch wave spectrum and Anderson localization length. Fig.~\ref{fig:Smatrix_bstructure} shows excellent agreement for $J/\mathrm{FSR} = 0.02$, representative of the experiments reported in Refs.~\cite{Hafezi13} [$\mathrm{FSR} \approx 1$ THz, $J \approx 20$ GHz]. The main discrepancies compared to the tight binding Hamiltonian are a small ($\approx 10\%$) reduction of the overall bandwidth, and a slight shift of the coupling asymmetries required to obtain the ``sawtooth-like'' flat bands: $\eta = 0.23\pi, 0.77\pi$. Thus, our use of a tight binding model in the main text is justified.

\section{Anderson localization length calculation} 	
In this Appendix, we discuss how localization length can be obtained analytically from the equation for the self -energy, \begin{widetext}
\begin{equation}
 	-\text{Im}\ \overline{\langle u_{+}(k)|\hat{\Sigma}(k,E)|u_{+}(k)\rangle}/\pi = \int dk' \overline{|\langle u_{+}(k)|\hat{V}|u_{+}(k')\rangle|^{2}} \delta(\omega(k) - \omega_{+}(k')) + \int dk'  \overline{|\langle u_{+}(k)|\hat{V}|u_{-}(k')\rangle|^{2}} \delta(\omega(k) - \omega_{-}(k')). 
 	\label{Selfenergy_full}
 \end{equation}
Contributions of each term are following: the first term describes intra-band scattering, while the second accounts for inter-band scattering. If the system is gapped, one can safely separate two terms. Now, we shall obtain the expression of $\hat\Sigma$ associated with different disorder symmetries. Given disorder $V \in [-\frac{W}{2},\frac{W}{2}]$, let us consider the second moment of disorder profile to calculate first order self energy. As mentioned in Eq.~\eqref{eq:disorder}, we allow three types for symmetries.
\subsection{Asymmetric disorder}
  Firstly, we take a look at the case when the disorder has no symmetry. Covariance of disorder has the form,
 	\begin{equation}
 		\overline{V_{i}V_{j}} = \frac{W^{2}}{12}\delta_{ij},
 	\end{equation}
 	where $\delta_{ij}$ is the Kronecker-delta function.
 \\\\1.\textit{ $\eta = \pi/4$, Dispersive band} -
 	One can derive the set of eigenstates for each dispersive band $|u_{D}\rangle$, and flat band $|u_{F}\rangle$ from Eq.~\eqref{Hamiltonian matrix} with $\eta = \pi/4$, Results are given
 	    {\footnotesize\begin{equation}
	\begin{aligned}
	&|u_{D}(k)\rangle = \frac{1}{\sqrt{(2\cos(k)+2\sqrt{2})^{2}-2\sin(2k)-4\sqrt{2}\sin k}}\left(2+\sqrt{2}\cos (k) - \sqrt{2}\sin (k) , \sqrt{2}+2\cos(k)\right), \quad \omega_{D}(k) = J(1+\sqrt{2}\cos k), \\
	&|u_{F}(k)\rangle = \frac{1}{\sqrt{(2\cos(k)+2\sqrt{2})^{2}-2\sin(2k)-4\sqrt{2}\sin k}}\left(2+\sqrt{2}\cos(k)+\sqrt{2}\sin(k), -(\sqrt{2}+2\cos(k))\right), \quad \omega_{F}(k) = -J. 
	\end{aligned}
	\label{eigenstate 0.25pi}
	\end{equation}}
 	In this case, we can only obtain the scattering time for dispersive band since group velocity of flat band is zero. In addition, as system exhibits gapped band profile, only the intra-band term in Eq.~\eqref{Selfenergy_full} contributes since only this part is nonzero. Eq.~\eqref{Selfenergy_full} is then 
 	 \begin{equation}
 	- \text{Im}\ \overline{\langle u_{D}(k)|\hat{\Sigma}(k,E)|u_{D}(k)\rangle}/\pi =  \left|\frac{d\omega(k')}{dk'}(k)\right|^{-1} \left(\int dk' \overline{|\langle u_{D}(k)|\hat{V}|u_{D}(k')\rangle|^{2}}\delta(k + k')\right). 
 	\label{selfenergy_dispersive}
 	\end{equation}
 	As $(\tau_{D}(k))^{-1} = -\text{Im}\overline{\langle \hat{\Sigma}(k)\rangle}_{D}/\pi$, we obtain the inverse scattering time $1/\tau(k)$ in this band,
 	 \begin{equation}
 	\begin{aligned}
 	\frac{1}{\tau_{D}(k)} &= |\sqrt{2}\sin(k)J|^{-1}\left(\overline{|\langle u_{D}(k)|\hat{V}|u_{D}(-k)\rangle|^{2}}\right), \\&= \frac{1}{\sqrt{2J^{2}-(\omega(k)-J)^{2}}}\left(\overline{\frac{V^{2}_{a}\left((\sqrt{2}\cos k + 2)^{2}-2\sin^{2}k \right)^{2}+V^{2}_{b}(2\cos k + \sqrt{2})^{4} + 2V_{a}V_{b} ....}{|(2\cos(k)+2\sqrt{2})^{4}-(4\sin k \cos k + 4\sqrt{2}\sin k)^{2}|}}\right),  
 	\\&= \frac{W^{2}}{24\sqrt{2J^{2}-(\omega(k)-J)^{2}}}\left(\frac{\left(\frac{\omega(k)}{J}\right)^{4}}{\left|\left(\frac{1}{\sqrt{2}}\left(\frac{\omega(k)}{J}+1\right)\right)^{2}\left(\frac{\omega(k)}{J}\right)^{2}\right|}\right) 
 	=  \frac{W^{2}}{12\sqrt{2J^{2}-(\omega(k)-J)^{2}}}\left(\frac{\omega(k)^{2}}{(\omega(k)+J)^{2}}\right).
 	\label{localization length 0.25pi}
 	\end{aligned}
 	\end{equation}
 	\end{widetext}
 	Localization length is thus
 	\begin{equation}
 	\xi(\omega) = 2v_{g}(\omega)\tau_{D}(\omega) = \frac{24(\omega+J)^{2}(2J^{2}-(\omega-J)^{2})}{W^{2}\omega^{2}}, 
 	\end{equation}
 	where $v_{g}(\omega) = |\frac{d\omega}{dk} (\omega)|$ and $(1-\sqrt{2})J \leq \omega \leq (1+\sqrt{2})J$. One can check $\xi$ diverges when $\omega = 0$ since the scattering time diverges! It results that $\omega = 0$ is immune to disorder under the Born approximation. It is equivalent result to spin-momentum locking point that we obtained in Sec. II. Meanwhile, $\xi$ vanishes at the band edges $\omega = (1\pm\sqrt{2})J$, because $v_{g}$ vanishes. Hence it shows a strong sensitivity to disorder. 
 	\\\\2.\textit{ $\eta = \pi/2$} - 
 	The dispersion is symmetric about zero detuning, $\omega_{+}(k) = -\omega_{-}(k)$, and bands do not overlap except band crossing point $k = \pm \pi$. Due to this crossing point, it looks like we should take the inter-band term in Eq.~\eqref{Selfenergy_full} into account. However, it turns out that in the vicinity of $\omega(k=\pm \pi) = 0$, first order perturbation theory breaks down and it is required higher order perturbation theory. In this first order approximation, we only consider the spectrum for nonzero detunings. Then, one can obtain the localization length from the intra-band term of Eq.~\eqref{Selfenergy_full}.
    From Eq.~\eqref{Hamiltonian matrix} with $\eta = \pi/2$, one can easily derive the set of eigenstates
    \begin{equation}
	\begin{aligned}
	 &
	 |u_{+}(k)\rangle = \frac{1}{\sqrt{2-2\sin(k/2)}}(1 - \sin(k/2) , \cos(k/2))^{T},  \\
		&|u_{-}(k)\rangle = \frac{1}{\sqrt{2+2\sin(k/2)}}(1 + \sin(k/2) , -\cos(k/2))^{T},
	\end{aligned}
	\end{equation}
	where $\omega_{\pm} = \pm 2J\cos(k/2)$.
 	In this calculation, we consider the positive band only due to symmetric profile. Inverse of scattering time is then
 	\begin{widetext}
 	    	\begin{equation}
	\begin{aligned}
	\frac{1}{\tau_{+}(k)} &= |J\sin(k/2)|^{-1}\left({|\langle u_{+}(k)|\hat{V}|u_{+}(-k)\rangle|^{2}}\right) =\frac{2}{\sqrt{4J^{2}-\omega^{2}}}\left({|\langle u_{+}(k)|\hat{V}|u_{+}(-k)\rangle|^{2}}\right), 
	 \\&=  \frac{2}{\sqrt{4J^{2}-\omega^{2}}}\left(\overline{\left|\frac{1}{\sqrt{2+2\sin(k/2)}}\frac{1}{\sqrt{2-2\sin(k/2)}} \left(V_{a}(1-\sin^{2}(k/2))+V_{b}(\cos^{2}(k/2))\right)\right|^{2}}\right), \\&= \frac{1}{2 \sqrt{4J^{2}-\omega^{2}}}\left(\overline{ V^{2}_{a} (1-\sin^{2}(k/2))+V^{2}_{b}(1-\sin^{2}(k/2)) + 2V_{a}V_{b}(1-\sin^{2}(k/2))}\right)
	 = \frac{W^{2}\omega^{2}}{48J^{2}\sqrt{4J^{2}-\omega^{2}}}.
	 \label{localization length 0.5pi}
	\end{aligned}	
	\end{equation}
 	\end{widetext}
	Hence Anderson localization length reads
	\begin{equation}
	\xi = \frac{48J^{2}(4J^{2}-\omega^{2})}{W^{2}\omega^{2}}, \quad (-2J \leq \omega \leq 2J). 
	\end{equation}
	One can observe that localization length diverges at $\omega = 0$, due to the spin-momentum locking of the eigenstates. Again, $\xi$ vanishes at band edges $\omega = \pm 2J$.
	\subsection{Locally correlated disorder}
	Now, we consider the case when two disorders in each sublattice are locally correlated. It means that $\overline{V_{i}V_{j}} \neq 0$. We consider two cases of correlated disorder: 1) $\overline{V_{i}V_{j}} = \frac{W^{2}}{12}$ (symmetric disorder), 2) $\overline{V_{i}V_{j}} = -\frac{W^{2}}{12}$ (anti-symmetric disorder). 
\\\\	1. \textit{$\eta = \pi/4$, Dispersive band} -
	From Eq.~\eqref{localization length 0.25pi}, we include the contribution from different sublattice correlation $\overline{V_{a}V_{b}}$. Additional contribution yields
	{\small\begin{equation}
	\begin{aligned}
	&\frac{1}{\tau_{D}(k)} 
	= \begin{cases}
 \frac{W^{2}}{6\sqrt{2J^{2}-(\omega-J)^{2}}}\left(\frac{\omega^{2}}{(\omega+J)^{2}}\right)\quad (\text{symmetric}), \\ 0 \qquad\quad\qquad\qquad\qquad(\text{anti-symmetric}).
	\end{cases}
	\end{aligned}
	\end{equation}}
	Localization length is then
	\begin{equation}
	\xi = \begin{cases}
	\frac{12(\omega+J)^{2}(2J^{2}-(\omega-J)^{2})}{W^{2}\omega^{2}}\quad (\text{symmetric}), \\ \infty \qquad\qquad\qquad\qquad(\text{anti-symmetric}),
	\end{cases}
	\end{equation}
	where $(1-\sqrt{2})J \leq \omega \leq (1+\sqrt{2})J$.
	\\\\2. \textit{$\eta = \pi/2$} -
	Like the previous case, from Eq.~\eqref{localization length 0.5pi}, we obtain
	\begin{equation}
	\begin{aligned}
	\frac{1}{\tau_{+}(k)} = \begin{cases}
	\frac{W^{2}\omega^{2}}{24J^{2}\sqrt{4J^{2}-\omega^{2}}} \quad (\text{symmetric}),\\ 0 \qquad\qquad(\text{anti-symmetric}).
	\end{cases}
	\end{aligned}
	\end{equation}
	Again, localization length is then
	\begin{equation}
	\xi = \begin{cases}
		\frac{24J^{2}(4J^{2}-\omega^{2})}{W^{2}\omega^{2}} \quad (\text{symmetric}),\\ \infty \qquad\qquad(\text{anti-symmetric}).
	\end{cases} \quad (-2J \leq \omega \leq 2J)
	\end{equation}

	Thus, we find that symmetric disorder reduces localization length by half, while anti-symmetric disorder leads to infinite localization length in the first order Born approximation.

\section{Transfer matrix method for Anderson localization length}

Here, we outline the transfer matrix method used to numerically obtain the Anderson localization length. For the sake of simplicity, let us begin with the Hamiltonian Eq.~\eqref{Hamiltonian} in the first quantization form via considering semi-classical field. In the presence of disorder, tight binding equation for the field amplitude $\psi = (a_{n},b_{n})$ reads
{\footnotesize \begin{equation}
	\begin{cases}
	\omega^{(a)}_{n}a_{n} = J\sin\eta b_{n} + \frac{J}{2}(e^{-i\eta}a_{n-1}+e^{i\eta}a_{n+1}+b_{n-1}+b_{n+1}),\\
	\omega^{(b)}_{n}b_{n}= J\sin\eta a_{n} + \frac{J}{2}(e^{i\eta}b_{n-1}+e^{-i\eta}b_{n+1}+a_{n-1}+a_{n+1}),
	\label{TB equation}
	\end{cases}
	\end{equation}}
where $\omega^{(r)}_{n} \coloneqq \omega - V^{(r)}_{n}$ $(r = a$ or $b$). Unfortunately, the transfer matrix is singular in this form~\cite{Dwivedi16}. The hopping matrices describing the coupling to neighbouring cells are not invertible. In other words, while we have two degrees of freedom per unit cell, there is only a single propagation channel between unit cells. To obtain a non-sigular transfer matrix, let us rewrite the Eq.~\eqref{TB equation} in a different basis. Define rotated amplitude basis $a'_{n} = e^{-i\eta}(a_{n}+b_{n})/2$ and $b'_{n} = (a_{n}-b_{n})/2$. Eq.~\eqref{TB equation} is then,
\begin{widetext}
	\begin{equation}
	\begin{cases}
	Je^{i\eta} a'_{n+1} = (\omega - V^{a}_{n} - Je^{i\eta}\sin\eta) a'_{n} + (\omega - V^{a}_{n} + Je^{i\eta}\sin\eta) b'_{n} - J\cos\eta a'_{n-1} + iJ\sin\eta b'_{n-1}, \\ Je^{i\eta} b'_{n+1} = (\omega - V^{b}_{n} - Je^{-i\eta}\sin\eta) a'_{n} + (\omega - V^{b}_{n} - Je^{-i\eta}\sin\eta) b'_{n} - J\cos\eta a'_{n-1} - iJ\sin\eta b'_{n-1}.
	\label{TB transformed equation}
	\end{cases}
	\end{equation}
	Here, we define  $r_{n} \coloneqq a'_{n+1}/a'_{n}$ and $q_{n} \coloneqq b'_{n}/a'_{n}$. By subtracting both equations with respect to $b'_{n}$, we obtain the equation for $r_{n}$,
		{\footnotesize\begin{equation}
		r_{n} = \frac{\left(2(V^{a}_{n}-\omega)(\omega - V^{b}_{n})+2\sin\eta + J[J\sin\eta(1+\cos 2\eta)-(V^{a}_{n}+V^{b}_{n}-2\omega)\cos\eta]r_{n-1}^{-1}+iJ\sin\eta[(V^{a}_{n}+V^{b}_{n}-2\omega) -J\sin 2\eta]\frac{q_{n-1}}{r_{n-1}}\right)}{Je^{-i\eta}(-iJ+V^{a}_{n}+e^{2i\eta}(iJ+V^{b}_{n}-\omega)-\omega)}.
	\end{equation}}
	In addition, subtraction with respect to $a_{n+1}$ yields the equation for $q_{n}$,
	\begin{equation}
	q_{n} = \frac{(V^{b}_{n}-V^{a}_{n})\cos\eta + i(V^{a}_{n}+V^{b}_{n}-2\omega)\sin\eta + J\sin 2 \eta \ r_{n-1}^{-1} + 2J\sin^{2}\eta\frac{q_{n-1}}{r_{n-1}}}{(V^{a}_{n}+V^{b}_{n}-2\omega)\cos\eta + (-2J-i(V^{a}_{n}-V^{b}_{n}))\sin\eta}.
	\end{equation}
	\end{widetext}

	One can iterate these two equations to obtain the localization length via calculating the ratio $r_{n}$ for the set of given initial conditions $(a_{0},b_{0})$, such that
	\begin{equation}
	\langle |r_{n}| \rangle \approx \exp(1/\xi).
	\end{equation}
	Where $\langle ... \rangle$ is the ensemble average. Strictly speaking, the map for $r_{n}$ has two eigenstates, but the growing one dominates. Hence we obtain the localization length $\xi$,
	\begin{equation}
	\xi^{-1} = \langle \log(|r_{n}|) \rangle. 
	\end{equation}
\end{appendix}

\bibliographystyle{apsrev4-1}
\bibliography{New}
\end{document}